\documentclass[12pt]{iopart}
\usepackage{hyperref,amssymb,graphicx,multirow,bm}
\pdfoutput=1
\newcommand\T{\rule{0pt}{3.1ex}}
\newcommand\B{\rule[-1.7ex]{0pt}{0pt}}

\begin{document}
\title[Classical big-bounce cosmology]{Classical big-bounce cosmology: dynamical analysis of a homogeneous and irrotational Weyssenhoff fluid}
\author{S D Brechet, M P Hobson, A N Lasenby}
\address{Astrophysics Group, Cavendish Laboratory, J.~J.~Thomson Avenue, Cambridge, CB3 0HE, UK}
\eads{\mailto{sdb41@mrao.cam.ac.uk}, \mailto{mph@mrao.cam.ac.uk}, \mailto{a.n.lasenby@mrao.cam.ac.uk}}
\begin{abstract}
A dynamical analysis of an effective homogeneous and irrotational Weyssenhoff fluid in general relativity is performed using the $1+3$ covariant approach that enables the dynamics of the fluid to be determined without assuming any particular form for the space-time metric. The spin contributions to the field equations produce a bounce that averts an initial singularity, provided that the spin density exceeds the rate of shear. At later times, when the spin contribution can be neglected, a Weyssenhoff fluid reduces to a standard cosmological fluid in general relativity. Numerical solutions for the time evolution of the generalised scale factor $R(t)$ in spatially-curved models are presented, some of which exhibit eternal oscillatory behaviour without any singularities. In spatially-flat models, analytical solutions for particular values of the equation-of-state parameter are derived. Although the scale factor of a Weyssenhoff fluid generically has a positive temporal curvature near a bounce (i.e. $\ddot{R}(t)>0$), it requires unreasonable fine tuning of the equation-of-state parameter to produce a sufficiently extended period of inflation to fit the current observational data.
\end{abstract}
\submitto{\CQG}
\maketitle

\section{Introduction}
The Einstein-Cartan (EC) theory of gravity is an extension of Einstein's theory of general relativity (GR) that includes the spin properties of matter and their influence on the geometrical structure of space-time ($\cite{Cartan:1922}$; see also $\cite{Kleinert:1989}$, $\cite{Kleinert:2008}$). In GR, the energy-momentum of the matter content is assumed to be the source of curvature of a Riemannian space-time manifold $V_4$. In the EC theory, the spin of the matter has been postulated, in addition, to be the source of torsion of a Riemann-Cartan space-time manifold $U_4$ $\cite{Hehl:1973}$. Weyssenhoff and Raabe $\cite{Weyssenhoff:1947}$ were the first to study the behaviour of perfect fluids with spin. Obukhov and Korotky extended their work in order to build cosmological models based on the EC theory $\cite{Obukhov:1987}$ and showed that by assuming the Frenkel condition$\footnote{Note that the Frenkel condition arises naturally when performing a rigorous variation of the action. It simply means that the spin pseudovector is spacelike in the fluid rest frame.}$ the theory may be described by an effective fluid in GR where the effective stress-energy momentum tensor contains some additional spin terms.

The aim of this publication is two-fold. First, we wish to investigate the possibility that the spin contributions for a Weyssenhoff fluid may avert an initial singularity, as first suggested by Trautman $\cite{Trautman:1973}$. Second, since any realistic cosmological model has to include an inflation phase to fit the current observational data, it is also of particular interest to see if the spin contributions are able to generate a dynamical model endowed with an early inflationary era, as first suggested by Gasperini $\cite{Gasperini:1986}$. Scalars fields can generate inflation, but they have not yet been observed. Therefore, it is of interest to examine possible alternatives, such as a Weyssenhoff fluid. In contrast to the approaches of Trautman $\cite{Trautman:1973}$ and Gasperini $\cite{Gasperini:1986}$, our use of the $1+3$ covariant formalism enables us to determine the dynamics of a Weyssenhoff fluid without assuming any particular form for the space-time metric. 

The study of the dynamics of a Weyssenhoff fluid in a $1+3$ covariant approach was initiated by Palle $\cite{Palle:1998}$. His work has been revised and extended in our previous publication $\cite{Brechet:2007a}$. The present paper builds on $\cite{Brechet:2007a}$ to extend the work carried out first by Trautman $\cite{Trautman:1973}$ in an isotropic space-time, and Kopczynski $\cite{Kopczynski:1973}$ and Stewart $\cite{Stewart:1973}$ in an anisotropic space-time. It also generalises the analysis of the inflationary behaviour of Weyssenhoff fluid models made by Gasperini $\cite{Gasperini:1986}$ to anisotropic space-times.

In our dynamical analysis, we choose to restrict our study to a spatially homogeneous and irrotational Weyssenhoff fluid. This particular choice, which implies a vanishing vorticity and peculiar acceleration, has been motivated by underlying fundamental physical reasons. For a vanishing vorticity, the fluid flow is hypersurface-orthogonal, which means that the instantaneous rest spaces defined at each space-time point should mesh together to form a set of 3-surfaces in space-time $\cite{Ellis:1971}$. These hypersurfaces, which are surfaces of simultaneity for all the fluid observers, define a global cosmic time coordinate determined by the fluid flow. Moreover, by assuming that any peculiar acceleration vanishes, the cosmic time is then uniquely defined. It is worth mentioning that the absence of vorticity is an involutive property, which means that if it is true initially then it will remain so at later times as shown by Ellis et al $\cite{Ellis:1998}$. Finally, the assumption that there is no vorticity on all scales implies that the fluid has no global rotation. This is in line with recent Bayesian MCMC analysis of WMAP data performed by Bridges et {\it al.} $\cite{Bridges:2006}$. Their work confirms that a physical Bianchi $\mathrm{VII_h}$ model, which has a non-vanishing vorticity, is statistically disfavored by the data.

It is worth pointing out that Szydlowski and Krawiec $\cite{Szydlowski:2004}$ have considered an isotropic and homogeneous cosmological model in which a Weyssenhoff fluid is proposed as a potential candidate to describe dark energy at late times. In a subsequent publication $\cite{Krawiec:2005}$, the authors showed that it is not disfavoured by SNIa data, but it may be in conflict with CMB and BBN
observational constraints. By contrast, in this paper, we consider the full evolutionary history of an, in general, anisotropic universe with a Weyssenhoff fluid as its matter source, concentrating in particular on the `early universe' behaviour when the spin terms are significant. Indeed, at late times, when the spin contributions can be neglected, the Weyssenhoff fluid reduces to a standard cosmological fluid. We thus allow for the presence of a non-zero cosmological constant, in accord with current observational constraints.

In $\Sref{Weyssenhoff fluid description}$, we give a concise description of a Weyssenhoff fluid using a 1+3 covariant approach outlined in {\it Appendix} A. The spatial symmetries and macroscopic spin averaging procedure are discussed in $\Sref{Spatial symmetries and macroscopic spin averaging}$. In $\Sref{Dymamics of a homogeneous and irrotational Weyssenhoff fluid}$, we establish the relevant dynamical relations for a homogeneous and irrotational Weyssenhoff fluid. In $\Sref{Geodesic singularity analysis}$, we perform a geodesic singularity analysis for such a fluid. In $\Sref{Dynamical evolution: general considerations}$, we analyse the fluid dynamics. The behaviour of the generalised scale factor $R(t)$ of such a fluid in a spatially-curved models is discussed in $\Sref{Quantitative dynamical evolution of spatially-curved models}$ and explicit analytical solutions in spatially-flat models are given in $\Sref{homogeneous solutions}$ . For the reader's convenience, certain main results obtained in our earlier work $\cite{Brechet:2007a}$ will be repeated in the case of a homogeneous and irrotational Weyssenhoff fluid in $\Sref{Weyssenhoff fluid description}$ and $\Sref{Dymamics of a homogeneous and irrotational Weyssenhoff fluid}$. In this paper, we use the $(+,-,-,-)$ signature. To express our results in the opposite signature used by Ellis $\cite{Ellis:1998}$, the correspondence between physical variables can be found in $\cite{Brechet:2007a}$.

\section{Weyssenhoff fluid description}
\label{Weyssenhoff fluid description}

\subsection{Weyssenhoff fluid phenomenology}

In the EC theory, the effect of the spin density tensor is locally to induce torsion in the structure of space-time. In holonomic coordinates, the torsion tensor ${T^{\lambda}}_{\mu\nu}$ is defined as the antisymmetric part of the affine connection ${{\tilde{\Gamma}}^{\lambda}}_{\ \mu\nu}$,
\begin{equation}
{T^{\lambda}}_{\mu\nu}={\tilde{\Gamma}^{\lambda}}_{\ [\mu\nu]}=
{\textstyle\frac{1}{2}}\left({\tilde{\Gamma}^{\lambda}}_{\
\mu\nu}-{\tilde{\Gamma}^{\lambda}}_{\ \nu\mu}\right)\
,\label{torsion def}
\end{equation}
which vanishes in GR since the connection is assumed to be symmetric in its two lower indices. Note that the tilde denotes an EC geometrical object to differentiate it from an effective GR object. 

The Weyssenhoff fluid is a continuous macroscopic medium which is characterised on microscopic scales by the spin of the matter fields. The spin density is described by an antisymmetric tensor,
\begin{equation}
S_{\mu\nu}=-S_{\nu\mu}\ ,\label{Spin density tens}
\end{equation}
which is the source of torsion,
\begin{equation}
{S^{\lambda}}_{\mu\nu}=u^{\lambda}S_{\mu\nu}\ .\label{Spin density
tens II}
\end{equation}
This fluid satisfies the Frenkel condition, which requires the intrinsic spin of a matter field to be spacelike in the rest frame of the fluid,
\begin{equation}
S_{\mu\nu}u^{\nu}=0\ ,\label{Frenkel}
\end{equation}
where $u^{\lambda}$ is the $4$-velocity of the fluid element. This condition implies an algebraic coupling between spin and torsion according to,
\begin{equation}
T^{\lambda}_{\ \mu\nu}=\kappa u^{\lambda}S_{\mu\nu}\
,\label{algebraic coupling}
\end{equation}
and arises naturally from a rigorous variation of the action as shown by $\cite{Obukhov:1987}$. Thus, the torsion contributions to the EC field equations are entirely described in terms of the spin density. It is also useful to introduce a spin-density scalar defined as,
\begin{eqnarray}
S^2\equiv{\textstyle\frac{1}{2}}S_{\mu\nu}S^{\mu\nu}\geq 0\ .\label{Spin density
scalar}
\end{eqnarray}

Obukhov and Korotky showed $\cite{Obukhov:1987}$ that for a perfect fluid the EC field equations reduce to effective GR Einstein field equations with additional spin terms, and a spin field equation.

The former are found to be,
\begin{equation}
R_{\mu\nu}-{\textstyle\frac{1}{2}}g_{\mu\nu}\mathcal{R}=\kappa T^{s}_{\mu\nu}\ ,
\label{Einstein ef eq}
\end{equation}
where the effective stress energy momentum tensor of the fluid is given by,
\begin{equation}
T^{s}_{\mu\nu}=(\rho_s+p_s)u_{\mu}u_{\nu}-p_sg_{\mu\nu}
-2\left(g^{\rho\lambda}+u^{\rho}u^{\lambda}\right)\nabla_{\rho}\left[u_{(\mu\vphantom)}S_{\vphantom(\nu)\lambda}\right]\ , \label{Ef stress en tensor}
\end{equation}
with effective energy density and pressure of the form,
\begin{eqnarray}
\eqalign{\rho_s=\rho-\kappa S^2+\kappa^{-1}\Lambda\ , \\
p_s=p-\kappa S^2-\kappa^{-1}\Lambda\ ,}
\label{energy density and pressure}
\end{eqnarray}
and the physical energy density and pressure satisfy the equation of state,
\begin{equation}
p=w\rho\ ,\label{equation of state}
\end{equation}
where  $\kappa=8\pi G$,
$\Lambda$ is the cosmological constant and $w$ the equation of state parameter.

The spin field equation is given by,
\begin{equation}
\nabla_{\lambda}\left(u^{\lambda}S_{\mu\nu}\right)
=2u^{\rho}u_{[\mu\vphantom]}\nabla_{|\lambda}\left(u^{\lambda}S_{\vphantom[\rho|\nu]}\right)\
. \label{Ef spin field equations}
\end{equation}

\subsection{Weyssenhoff fluid description in a 1+3 covariant formalism}

The $1+3$ covariant formalism outlined in {\it Appendix} A can now be used to perform a more transparent analysis of the Weyssenhoff fluid dynamics. Using a $1+3$ covariant approach in $\cite{Brechet:2007a}$, we found that the symmetric stress-energy momentum tensor $\eref{Ef stress en tensor}$ can be recast as, 
\begin{eqnarray}
T_{\mu\nu}^{s}=\rho_su_{\mu}u_{\nu}-p_sh_{\mu\nu}-2{\sigma_{(\mu\vphantom)}}^{\lambda}S_{\vphantom(\nu)\lambda}-2u_{(\mu\vphantom)}D^{\lambda}S_{\vphantom(\nu)\lambda}\
, \label{stress energy mom 1+3} 
\end{eqnarray}
where $h_{\mu\nu}$ is the induced metric on the spatial hypersurface, $\sigma_{\mu\nu}$ is the rate of shear tensor and $D^{\lambda}$ is the spatially projected covariant derivative defined in {\it Appendix} A.

Similarly, the spin field equation $\eref{Ef spin field equations}$ reduces to, \begin{eqnarray}
\dot{S}_{\mu\nu}+\Theta S_{\mu\nu}=2u^{\rho}u_{[\mu\vphantom]}\dot{S}_{\vphantom[
|\rho|\nu]}\ , \label{Antisym field equa} 
\end{eqnarray}
where $\Theta=D^{\lambda}u_{\lambda}$ is the expansion rate.

\section{Spatial symmetries and macroscopic spin averaging}
\label{Spatial symmetries and macroscopic spin averaging}

Although much of our following discussion will concern cosmological models that are anisotropic, it is of interest to consider the status of a Weyssenhoff fluid as a matter source for homogeneous and isotropic models.

\subsection{Spatial symmetries}

To be a suitable candidate for the matter content of such a cosmological model, a Weyssenhoff fluid has to be compatible with the Cosmological Principle. In mathematical terms, a four-dimensional space-time manifold satisfying this principle is foliated by three dimensional spatial hypersurfaces, which are maximally symmetric and thus invariant under the action of translations and rotations.

Although a Weyssenhoff fluid can be expressed as an effective GR fluid, the dynamical nature of such a fluid is rooted in the EC theory. Thus, the dynamics of such a fluid is determined by the translational and the rotational fields, which are respectively the metric $g_{\mu\nu}$ and the torsion ${T^{\lambda}}_{\mu\nu}$. The symmetries require the dynamical fields to be invariant under the action of an infinitesimal isometry. Hence, the Lie derivatives of the dynamical fields have to vanish according to,
\begin{eqnarray}
\mathcal{L}_{\xi}g_{\mu\nu}=0\ ,\label{Lie deriv metric}\\
\mathcal{L}_{\xi}{T^{\lambda}}_{\mu\nu}=0\ .\label{Lie deriv torsion}
\end{eqnarray} 
where $\xi^{\mu}$ are the Killing vectors generating the spatial isometries. A maximally symmetric spatial hypersurface admits 6 Killings vectors $\cite{Weinberg:1972}$. The 3 Killing vectors $\xi^{\mu}$ generating the infinitesimal translations are related to homogeneity and the 3 Killing pseudo-vectors $\chi^{\mu}$ generating the infinitesimal rotations are related to isotropy. They satisfy,
\begin{eqnarray}
\xi^{\mu}={h^{\mu}}_{\rho}\xi^{\rho}\ ,\label{Lie Isotropy}\\
\chi^{\mu}=\epsilon^{\mu\rho\sigma}\tilde{D}_{[\rho\vphantom]}\xi_{\vphantom[\sigma]}\ ,\label{Lie Homogeneity}
\end{eqnarray}
where $\epsilon^{\mu\rho\sigma}$ is three-dimensional Levi-Civita tensor.

For a cosmological fluid based on the EC theory, such as a Weyssenhoff fluid, we can consider two different forms of the Cosmological Principle:
\begin{enumerate}
\item the Strong Cosmological Principle (SCP), where the Lie derivatives of the metric $\eref{Lie deriv metric}$ and of the torsion $\eref{Lie deriv torsion}$ have to vanish; and
\item the Weak Cosmological Principle (WCP), where only the Lie derivatives of the metric $\eref{Lie deriv metric}$ have to vanish and no restriction is imposed on the torsion.	
\end{enumerate} 

The translational Killing equation resulting from the symmetries imposed on the metric  $\eref{Lie deriv metric}$ yields,
\begin{equation}
\tilde{D}_{(\mu\vphantom)}\xi_{\vphantom(\nu)}=D_{(\mu\vphantom)}\xi_{\vphantom(\nu)}=0\ ,\label{Killing metric}
\end{equation}
which is a well-know result obtained in GR. Hence, the WCP is identical to the GR Cosmological Principle, which implies that the space-time geometry is described in terms of an FRW metric.

Using the translational Killing equation $\eref{Killing metric}$, the rotational Killing equation resulting from the symmetries imposed on the metric $\eref{Lie deriv torsion}$ is found to be, 
\begin{equation}
(\tilde{D}_{\rho}{T^{\lambda}}_{\mu\nu})\xi^{\rho}=
(h^{\sigma\lambda}{T^{\rho}}_{\mu\nu}+{h^{\sigma}}_{\mu}{T^{\lambda\rho}}_{\nu}+{h^{\sigma}}_{\nu}T_{\ \; \mu}^{\lambda\ \,\rho})\tilde{D}_{[\rho\vphantom]}\xi_{\vphantom[\sigma]}\ .\label{Killing torsion}
\end{equation}

For any maximally symmetric space $\cite{Weinberg:1972}$, we can choose respectively a Killing vector $\xi^{\mu}$ to vanish at a given point $P$, and independently, a Killing pseudo-vector $\chi^{\mu}$ to vanish at a given point $Q$ according to,
\begin{eqnarray}
\xi^{\mu}(P)=0\ ,\label{Lie Iso}\\
\chi^{\mu}(Q)=0\ .\label{Lie Homo}
\end{eqnarray}
Hence, the homogeneity and isotropy can be considered separately.

By imposing the homogeneity condition $\eref{Lie Homo}$ on the rotational Killing equation $\eref{Killing torsion}$, the spatial covariant derivative of the torsion tensor has to vanish according to,
\begin{equation}
\tilde{D}_{\rho}T_{\lambda\mu\nu}=0\ .	
\end{equation}
Hence, torsion can only be a function of cosmic time $t$,
\begin{equation}
T_{\lambda\mu\nu}\equiv T_{\lambda\mu\nu}(t)\ .\label{Homo requirement}
\end{equation}

By imposing the isotropy condition $\eref{Lie Iso}$ on the rotational Killing equation $\eref{Killing torsion}$, the torsion tensor has to satisfy the constraint,
\begin{equation}
{h^{[\rho\vphantom]}}_{\tau}{h^{\vphantom[\sigma]}}_{\lambda}{T^{\tau}}_{\mu\nu}+
{h^{[\rho\vphantom]}}_{\tau}{h^{\vphantom[\sigma]}}_{\mu}T^{\ \tau}_{\lambda\ \,\nu}+
{h^{[\rho\vphantom]}}_{\tau}{h^{\vphantom[\sigma]}}_{\nu}{T_{\lambda\mu}}^{\tau}=0\ .\label{Iso requirement}
\end{equation}

As shown explicitly in a theorem established by Tsamparilis $\cite{Tsamparilis:1979}$ and mentioned subsequently by Boehmer $\cite{Boehmer:2006}$, the homogeneity $\eref{Homo requirement}$ and isotropy $\eref{Iso requirement}$ constraints taken together put severe restrictions on the torsion tensor. The only non-vanishing components are found to be,
\begin{eqnarray}
T_{\lambda\mu\nu}={h_{\lambda}}^{\alpha}{h_{\mu}}^{\beta}{h_{\nu}}^{\gamma}T_{[\alpha\beta\gamma]}=f(t)\epsilon_{\lambda\mu\nu}\ ,\label{totally antisym torsion}\\
{T^{\rho}}_{\mu\rho}=u_{\mu}u^{\gamma}{h_{\alpha}}^{\rho}{h_{\rho}}^{\beta}{T^{\alpha}}_{\gamma\beta}={\textstyle\frac{1}{3}}u_{\mu}u^{\gamma}{^{*}T_{\gamma}}\ ,\label{trace index torsion}
\end{eqnarray}
where $f(t)$ is a scalar function of cosmic time $t$, $\rho$ is a fixed index and $^{*}T_{\gamma}$ is the spatial trace of the torsion tensor defined as
\begin{eqnarray}
^{*}T_{\gamma}\equiv h^{\alpha\beta}T_{\alpha\gamma\beta}\ .\label{torsion trace}
\end{eqnarray}
We now discuss the application of this general framework to a Weyssenhoff fluid.

\subsection{Weyssenhoff fluid with macroscopic spin averaging}
The algebraic coupling between the spin density and torsion tensors $\eref{algebraic coupling}$ shows that the spin density $S_{\mu\nu}$ of a Weyssenhoff fluid can be related to the torsion as,
\begin{equation}
S_{\mu\nu}=u^{\alpha}{h_{\mu}}^{\beta}{h_{\nu}}^{\gamma}\kappa^{-1}T_{\alpha\beta\gamma}\ .\label{spin density torsion}
\end{equation}
By substituting the non-vanishing components of the torsion $\eref{totally antisym torsion}$ and $\eref{trace index torsion}$ satisfying the SCP into the expression for the spin density of a Weyssenhoff fluid $\eref{spin density torsion}$, it is straightforward to show that the spin density tensor has to vanish,
\begin{equation}
S_{\mu\nu}=0\ .\label{Vanish Spin Tensor}
\end{equation}

Thus, Tsamparilis claims that a Weyssenhoff fluid is incompatible with the SCP $\cite{Tsamparilis:1979}$. This conclusion would hold if all the dynamical contributions of the spin density were second rank tensors of the form $S_{\mu\nu}$. However, this is not the case since the dynamics contains spin density squared scalar terms. These scalar terms are invariant under spatial isometries like rotations and translations. Hence, they do satisfy the SCP. 

In order for the Weyssenhoff fluid to be compatible with the SCP, the spin density tensorial terms have to vanish leaving the scalar terms unaffected. This can be achieved by making the reasonable physical assumption that, locally, macroscopic spin averaging leads to a vanishing expectation value for the spin density tensor according to,
\begin{eqnarray}
\left<S_{\mu\nu}\right>=0\ .\label{spin averaging}	
\end{eqnarray}
However, this macroscopic spin averaging does not lead to a vanishing expectation value for the spin density squared scalar since this term is a variance term,
\begin{eqnarray}
\left<S^2\right>={\textstyle\frac{1}{2}}\left<S_{\mu\nu}S^{\mu\nu}\right>\neq0\ .\label{spin non-averaging}	
\end{eqnarray}
 
The macroscopic spatial averaging of the spin density was performed in an isotropic case by Gasperini $\cite{Gasperini:1986}$. It can be extended to an anisotropic case provided that on small macroscopic scales the spin density pseudo-vectors are assumed to be randomly oriented. 

By considering a Weyssenhoff fluid in the absence of any peculiar acceleration and by performing a macroscopic spin averaging, we indirectly require the fluid to be homogeneous. This follows from the fact that, in this case, the conservation law of momentum leads to a vanishing spatial derivative of the pressure and energy density. This will be explicitly shown in $\Sref{Conservation laws}$, and can also be derived from the corresponding dynamical equation for an inhomogeneous Weyssenhoff fluid presented in our previous work $\cite{Brechet:2007a}$.

Note that even in the absence of a macroscopic spin averaging, the Weyssenhoff fluid is still compatible with the WCP, which we discuss further in $\Sref{Comparision with Lu and Cheng}$. It is worth mentioning that there is no observational evidence so far which would suggest that we should impose the SCP even though the mathematical symmetries make such a principle mathematically appealing. A true test of whether this principle is applicable would be the demonstration of physically observable differences between this case and the WCP.

\section{Dynamics of a homogeneous and irrotational Weyssenhoff fluid}
\label{Dymamics of a homogeneous and irrotational Weyssenhoff fluid}

The dynamics of a Weyssenhoff fluid with no peculiar acceleration is entirely determined by the symmetric and spin field equations, $\eref{Einstein ef eq}$, $\eref{stress energy mom 1+3}$ and $\eref{Antisym field equa}$ respectively. The former can be used to determine the Ricci identities and the energy conservation law. The latter simply expresses spin propagation.

One important consequence of the spatial averaging of the spin density is that the stress-energy momentum tensor $\eref{stress energy mom 1+3}$ reduces to an elegant expression given by
\begin{eqnarray}
T_{\mu\nu}^{s}=\rho_su_{\mu}u_{\nu}-p_sh_{\mu\nu}\
, \label{stress energy mom 1+3 isotropic}
\end{eqnarray}
where the only spin contributions affecting the dynamics are the negative spin squared variance terms entering the definition of the effective energy density and pressure $\eref{energy density and pressure}$, as expected. These spin squared intrinsic interaction terms $S^2$ are a key feature that distinguishes a Weyssenhoff fluid from a perfect fluid in GR and lead to interesting properties we discuss below.

We have to be careful when performing the macroscopic spin averaging on the dynamical equations. The Ricci identities and conservation laws can be entirely determined from the stress-energy momentum tensor $\eref{stress energy mom 1+3 isotropic}$. As we have shown, it is perfectly legitimate to perform a macroscopic spin averaging on the stress-energy momentum tensor before obtaining explicitly the dynamical equations. However, this is not the case for the spin field equations $\eref{Antisym field equa}$. Performing the macroscopic spin averaging at this stage would make these field equations vanish. To be consistent, we first have to determine the dynamical equations and express them in terms of the spin density scalar before performing the spin averaging.  

\subsection{Ricci identities}

The Ricci identities can firstly be applied to the whole space-time and secondly to the orthogonal 3-space. They yield respectively,
\begin{eqnarray}
2\nabla_{[\mu\vphantom]}\nabla_{\vphantom[\nu]}u_{\rho}=R_{[\mu\nu]\rho
}^{\ \ \ \ \ \lambda}u_{\lambda}\
,\label{Ricci identities}\\
2D_{[\mu\vphantom]}D_{\vphantom[\nu]}v_{\rho}={\vphantom{a}^{*}R_{[\mu\nu]\rho}}^{\lambda}v_{\lambda}\
,\label{Projected Ricci}
\end{eqnarray}
where the spatial vectors $v^{\mu}$ are orthogonal to the worldline, i.e. $v^{\mu}u_{\mu}=0$, and the 3-space Riemann tensor ${\vphantom{a}^{*}R_{\mu\nu\rho\lambda}}$ is related to the Riemann
tensor ${R_{\mu\nu\rho\lambda}}$ by
\begin{eqnarray}
{\vphantom{a}^{*}R_{\mu\nu\rho\lambda}}=
{h^{\alpha}}_{\mu}{h^{\beta}}_{\nu}{h^{\gamma}}_{\rho}{h^{\delta}}_{\lambda}R_{\alpha\beta\gamma\delta}
+\Theta_{\mu\rho}\Theta_{\nu\lambda}-\Theta_{\mu\lambda}\Theta_{\nu\rho}\
.\label{Riemann tensor relations} 
\end{eqnarray}
The Riemann tensor can be decomposed according to its symmetries as $\cite{Hawking:1966}$,
\begin{equation}
R^{\rho\mu}_{\ \ \ \nu\lambda}=C^{\rho\mu}_{\ \ \ \nu\lambda} -
\delta^{\rho}_{\ [\lambda\vphantom]}R^{\mu}_{\ \vphantom[\nu]} -
\delta^{\mu}_{\ [\nu\vphantom]}R^{\rho}_{\ \vphantom[\lambda]} -
{\textstyle\frac{1}{3}}\mathcal{R}\delta^{\rho}_{\ [\nu\vphantom]}\delta^{\
\mu}_{\ \ \vphantom[\lambda]}\ ,\label{Riemann tens}
\end{equation}
where $C^{\rho\mu}_{\ \ \ \nu\lambda}$ is the trace-free Weyl tensor, which, in turn, can be split into an `electric' and a `magnetic' part $\cite{Hawking:1966}$ according to,
\begin{eqnarray}
E_{\mu\nu} = C_{\mu\rho\nu\sigma}u^{\rho}u^{\sigma}\ ,\label{Elec}\\
H_{\mu\nu} = {\textstyle\frac{1}{2}}\eta_{\mu\sigma\lambda}C^{\sigma\lambda}_{\ \ \
\nu\rho}u^{\rho}\ .\label{Magn}
\end{eqnarray}
The Ricci tensor $R_{\mu\nu}$ is then simply obtained by substituting the expression $\eref{stress energy mom 1+3 isotropic}$ for the effective stress energy momentum tensor $T_{\mu\nu}^{s}$ into the Einstein field equations $\eref{Einstein ef eq}$,
\begin{eqnarray}
R_{\mu\nu}={\textstyle\frac{\kappa}{2}}\left(\rho_s+3p_s\right)u_{\mu}u_{\nu}
-{\textstyle\frac{\kappa}{2}}\left(\rho_s-p_s\right)h_{\mu\nu}\ .\label{Ef Ricci tensor}
\end{eqnarray}
The Riemann tensor $R_{\rho\mu\nu\lambda}$ can now be recast in terms of the Ricci tensor $\eref{Ef Ricci tensor}$, the electric $\eref{Elec}$ and magnetic $\eref{Magn}$ parts of the Weyl tensor according to the decomposition $\eref{Riemann tens}$ in the following way,  
\begin{eqnarray}
\eqalign{\fl \qquad R^{\rho\mu}_{\ \ \nu\lambda}\ = &{\textstyle\frac{2}{3}}\kappa\left(\rho_s +
	3p_s\right)h^{[\rho\vphantom]}_{\ \
	[\nu\vphantom]}u^{\vphantom[\mu]}u_{\vphantom[\lambda]}
	- {\textstyle\frac{2}{3}}\kappa\rho_s h^{[\rho\vphantom]}_{\ \ [\nu\vphantom]}h^{\vphantom[\mu]}_{\ \ \vphantom[\lambda]}\\
	&+4u^{[\rho\vphantom]}u_{[\nu\vphantom]}E^{\vphantom[\mu]}_{\ \
	\vphantom[\lambda]} - 4h^{[\rho\vphantom]}_{\ \
	[\nu\vphantom]}E^{\vphantom[\mu]}_{\ \ \vphantom[\lambda]}
	+2\eta^{\rho\mu\sigma}u_{[\nu\vphantom]}H_{\vphantom[\lambda]\sigma}
	+2\eta_{\nu\lambda\sigma}u^{[\rho\vphantom]}H^{\vphantom[\mu]\sigma}\
	.}\label{Riemann tensor}
\end{eqnarray}
It follows from the relation $\eref{Riemann tensor relations}$ that the Riemann tensor on the spatial 3-space ${\vphantom{a}^{*}R^{\rho\mu}_{\ \ \ \nu\lambda}}$ becomes,
\begin{eqnarray}
	\eqalign{
	{\vphantom{a}^{*}R^{\rho\mu}_{\ \ \ \nu\lambda}}= &-{\textstyle\frac{2}{3}}\kappa{h^{\rho}}_{[\nu\vphantom]}{h^{\mu}}_{\vphantom[\lambda]}
	\rho_s-4{h^{\rho}}_{[\nu\vphantom]}{E^{\mu}}_{\vphantom[\lambda]}+2{\Theta^{\rho}}_{[\nu\vphantom]}{\Theta^{\mu}}_{\vphantom[\lambda]}\ .}\label{3 Riemann tensor}
\end{eqnarray}

The information contained in the Ricci identities $\eref{Ricci identities}-\eref{Projected Ricci}$ can now be extracted by projecting them on different hypersurfaces using the decomposition of the corresponding Riemann tensors $\eref{Riemann tensor}-\eref{3 Riemann tensor}$ and following the same procedure as in our previous publication $\cite{Brechet:2007a}$.

The Ricci identities applied to the whole space-time yield respectively the Raychaudhuri equation and the rate of shear propagation equation,
\begin{eqnarray}
\dot{\Theta}=-{\textstyle\frac{1}{3}}\Theta^2-2\sigma^2
-{\textstyle\frac{\kappa}{2}}\left(\rho_s + 3p_s\right)\ ,\label{Raychaudhuri eq}\\
\dot{\sigma}_{\langle\mu\nu\rangle}=-{\textstyle\frac{2}{3}}\Theta\,\sigma_{\mu\nu}
-\sigma_{\langle\mu}^{\ \ \lambda}\sigma_{\nu\rangle\lambda}-E_{\mu\nu}
\ .\label{Rate shear prop eq}
\end{eqnarray}

The Ricci identities applied to the spatial 3-space express the spatial curvature. Their contractions yield the spatial Ricci tensor $^{*}\mathcal{R}_{\mu\nu}$ and scalar $^{*}\mathcal{R}$ respectively,
\begin{eqnarray}
^{*}R_{\mu\nu}=\dot{\sigma}_{\langle\mu\nu\rangle}+\Theta\,\sigma_{\mu\nu}
+{\textstyle\frac{1}{3}}{h_{\mu\nu}}^{*}\mathcal{R}\ ,\label{Ricci tensor eq}\\
^{*}\mathcal{R}={\textstyle\frac{2}{3}}\Theta^2-2\kappa\rho_s-2\sigma^2\
.\label{Generalised Friedmann eq}
\end{eqnarray}
The above expression for the curvature scalar $\eref{Generalised Friedmann eq}$ is a generalisation of the Friedmann equation.

One must take particular care when deducing the time evolution of the rate of shear from the rate of shear propagation equation $\eref{Rate shear prop eq}$. This is due to the fact that the rate of shear coupling term $\sigma_{\langle\mu}^{\ \ \lambda}\sigma_{\nu\rangle\lambda}$ and the tidal force term $E_{\mu\nu}$ can not simply be neglected. A better route is to deduce the rate of shear evolution equation from the spatial Ricci curvature tensor $^{*}R_{\mu\nu}$ as shown explicitly by Ellis $\cite{Ellis:1973}$ and outlined below.

A homogeneous Weyssenhoff fluid satisfies the spatial curvature identity,
\begin{equation}
^{*}R_{\mu\nu}-{\textstyle\frac{1}{3}}{h_{\mu\nu}}^{*}\mathcal{R}=0\ .\label{Spatial curvature identity}
\end{equation}
Hence, by substituting this identity $\eref{Spatial curvature identity}$ into the expression for the spatial Ricci tensor $\eref{Ricci tensor eq}$, the propagation equation for the rate of shear is found to be,
\begin{equation}
\dot{\sigma}_{\langle\mu\nu\rangle}=-\Theta\,\sigma_{\mu\nu}\ .\label{Shear prop I}
\end{equation}
This tensorial expression $\eref{Shear prop I}$ can be recast in terms of a scalar relation involving the rate of shear scalar $\sigma$ according to,
\begin{equation}
\dot{\sigma}=-\Theta\,\sigma\ .\label{Shear prop II}
\end{equation}

\subsection{Conservation laws}
\label{Conservation laws}

The effective energy conservation and momentum conservation laws are obtained by projecting the conservation equation of the effective stress energy momentum tensor $\eref{stress energy mom 1+3 isotropic}$,

\begin{equation}
\nabla^{\mu}\left(R_{\mu\nu}+{\textstyle\frac{1}{2}}g_{\mu\nu}\mathcal{R}\right)=
\kappa\nabla^{\mu}T^{s}_{\mu\nu}=0\ ,\label{Twice Bianchi}
\end{equation}
respectively along the worldline $u^{\nu}$ and on the orthogonal spatial hypersurface $h_{\mu\nu}$ according to,
\begin{eqnarray}
\dot{\rho_s} = -\Theta\,\left(\rho_s+p_s\right)\ ,\label{Eff En cons
eq}\\
D_{\mu}p_s=0\ .\label{Mom cons eq}
\end{eqnarray}
It is worth mentioning that the momentum conservation law $\eref{Mom cons eq}$ expresses the homogeneity of the Weyssenhoff fluid. This is due to the fact that according to this law, the energy density, the pressure and the spin density of the fluid have to be a function of cosmic time only. Hence, the torsion tensor has also to be a function of cosmic time only, which is the homogeneity requirement $\eref{Homo requirement}$. This is only the case for a Weyssenhoff fluid with no peculiar acceleration on which a macroscopic spin averaging has been performed, as otherwise the momentum conservation law $\eref{Mom cons eq}$ would contain additional terms.

\subsection{Spin propagation relation}

The spin conversation law results from twice projecting the antisymmetric field equations $\eref{Antisym field equa}$ onto the hypersurface orthogonal to the worldline,
\begin{equation}
{\left(S_{\mu\nu}\right)}^{\dot{\vphantom a}}_{\bot}=-\Theta\,S_{\mu\nu}\ .\label{Spin cons vector eq}
\end{equation}
This tensorial expression $\eref{Spin cons vector eq}$ can be recast in terms of a scalar relation involving the spin-density scalar $S$ in $\eref{Spin density
scalar}$ according to,
\begin{equation}
\dot{S}=-\Theta\,S\ .\label{Spin cons eq}
\end{equation}
This expression implies that the spin density is inversely proportional to the volume of the fluid. Note that although the tensorial expression $\eref{Spin cons vector eq}$ vanishes due to the macroscopic spin averaging $\eref{spin averaging}$, the scalar expression $\eref{Spin cons eq}$ still applies because it is related to the spin variance $\eref{spin non-averaging}$.

The effective energy conservation equation $\eref{Eff En cons eq}$ can now be recast in terms of the true (i.e. not effective) energy density and pressure of the fluid by substituting the spin propagation equation $\eref{Spin cons eq}$,
\begin{equation}
\dot{\rho} = -\Theta\,\left(\rho+p\right)\ .\label{En cons eq}
\end{equation}

\subsection{Comparision with previous results}
\label{Comparision with Lu and Cheng}

Let us compare our results with the conclusions reached by Lu and Cheng $\cite{Lu:1995}$ for an isotropic Weyssenhoff fluid without any macroscopic spin averaging as shown in {\it Appendix A} of their publication. 

In an isotropic space-time, the dynamics of a Weyssenhoff fluid, without a macroscopic spin averaging, is greatly simplified as we now briefly explain. The projection of the effective Einstein field equations $\eref{Einstein ef eq}$ along the worldline and on the orthogonal spatial hypersurfaces, yields the following constraint,
\begin{equation}
u^{\mu}{h_{\lambda}}^{\nu}T^s_{\mu\nu}=0\ .\label{projection constraint}
\end{equation}
It arises from the fact that, in an isotropic case, the time-space components of the Ricci tensor vanish. From the expression for the stress-energy momentum tensor $\eref{stress energy mom 1+3}$, it is clear that the constraint $\eref{projection constraint}$ implies a vanishing spin divergence,
\begin{equation}
D^{\lambda}S_{\mu\lambda}=0\ .\label{spin divergence}
\end{equation}
Moreover, the isotropy constraint implies a vanishing rate of shear (i.e. $\sigma=0$). Thus, in this case, the effective stress energy momentum tensor without the macroscopic spin averaging  $\eref{stress energy mom 1+3}$ reduces to the elegant expression $\eref{stress energy mom 1+3 isotropic}$ obtained by performing the macroscopic spin averaging. 

Hence, for a Weyssenhoff fluid and isotropic space-time, our results can be compared to those of Lu and Cheng $\cite{Lu:1995}$. The results of our analysis do not agree with the conclusions outlined in $\cite{Lu:1995}$. First, they argue that the isotropic Friedmann equation implies that the spin density has to be a function of time only, with which we agree. Then, they claim that this stands in contradiction with the fact that the spin density has also to be a function of space in order to satisfy the projection constraint $\eref{projection constraint}$, which we dispute. The projection constraint simply implies a vanishing orthogonal projection of the spin divergence on the spatial hypersurface $\eref{spin divergence}$, which is perfectly compatible with the spin density being a function of time only. Hence, contrary to their claim, a Weyssenhoff fluid model seems to be perfectly consistent with an isotropic space-time (i.e. obeying the WCP), even without spin averaging.

\section{Geodesic singularity analysis}
\label{Geodesic singularity analysis}

For a homogeneous and irrotational Weyssenhoff fluid satisfying the macroscopic spin averaging condition, the fluid congruence is geodesic. To study the behaviour of such a fluid congruence near a singularity, we use the 1+3 covariant formalism, which applies on local as well as on global scales for a homogeneous fluid model. 

In order for singularities in the timelike geodesic congruence to occur, the Raychaudhuri equation $\eref{Raychaudhuri eq}$ has to satisfy the condition,
\begin{equation}
\dot{\Theta}+\frac{1}{3}\Theta<0\ ,\label{Ray cond}
\end{equation}
near the singularity, as we now explain. First, we recast the singularity condition $\eref{Ray cond}$ in terms of the inverse expansion rate $\Theta^{-1}$ as,
\begin{equation}
\frac{d}{dt}\left(\Theta^{-1}\right)>\frac{1}{3}\ ,\label{Ray cond II}
\end{equation}
After integrating with respect to cosmic time $t$, we find,
\begin{equation}
\Theta^{-1}(t)>\Theta^{-1}_{*}+\frac{1}{3}\left(t-t_{*}\right)\ ,\label{Ray cond III}
\end{equation}
where $\Theta^{-1}_{*}\equiv\Theta^{-1}\left(t_{*}\right)$ and $t=t_*$ is some arbitrary cosmic time near the singularity. Thus, if $\Theta^{-1}_{*}>0$ ($\Theta^{-1}_{*}<0$), the model describes a fluid evolving on a spatially expanding (collapsing) hypersurface at $t=t_{*}$. According to the integrated singularity condition $\eref{Ray cond III}$, $\Theta^{-1}\left(t\right)$ must vanish within a finite past (future) time interval $|t-t^{*}|<3|\Theta^{-1}_{*}|$ with respect to $t=t_{*}$. Thus a geodesic singularity, defined by $\Theta^{-1}(\hat{t})=0$, occurs at $t=\hat{t}$ .

The homogeneity requirement allows us to define $\--$ up to a constant factor $\--$ a generalised scale factor $R$ according to,
\begin{eqnarray}
\Theta\equiv 3\frac{\dot{R}}{R}\ .\label{scaling} 
\end{eqnarray}
In a $1+3$ covariant approach, $R$ is generally a locally defined variable. If the model is homogeneous, however, $R$ can be globally defined and interpreted as a cosmological scale factor.

The singularity condition can now be recast in terms of the scale factor $R$ and reduces to,
\begin{equation}
\frac{\ddot{R}}{R}<0\ .
\label{SE}
\end{equation}
One must also require the scale factor to obey the consistency condition, which requires the expansion rate squared to be positively defined at all times according to, 
\begin{equation}
\left(\frac{\dot{R}}{R}\right)^2>0\ .
\label{WE}
\end{equation}

To determine explicitly these two conditions $\eref{SE}$-$\eref{WE}$, the Friedmann $\eref{Generalised Friedmann eq}$ and Raychaudhuri $\eref{Raychaudhuri eq}$ equations are recast in terms of the scale factor, using respectively the expressions for the Ricci $\eref{Ef Ricci tensor}$ and stress-energy-momentum tensor $\eref{stress energy mom 1+3 isotropic}$ as,
\begin{eqnarray}
\left(\frac{\dot{R}}{R}\right)^2=\frac{1}{3}\left(T_{\mu\nu}u^{\mu}u^{\nu}+\frac{^{*}\mathcal{R}}{2}+\sigma^2\right)\ ,\label{Friedmann sing}\\
\frac{\ddot{R}}{R}=-\frac{1}{3}\left(R_{\mu\nu}u^{\mu}u^{\nu}+2\sigma^2\right)\ .
\label{Ricci tens eq}
\end{eqnarray}
Using the Friedmann $\eref{Friedmann sing}$ and the Raychaudhuri $\eref{Ricci tens eq}$ equations, the consistency $\eref{WE}$ and singularity $\eref{SE}$ conditions can respectively be explicitly expressed as,
\begin{eqnarray}
\frac{S^2-\kappa^{-2}\left(\sigma^2+\Lambda+{\textstyle\frac{1}{2}}^{*}\mathcal{R}\right)}{\kappa^{-1}\rho}<1\ ,\label{WE II}\\
\frac{S^2-\kappa^{-2}\left(\sigma^2-{\textstyle\frac{1}{2}}\Lambda\right)}{\kappa^{-1}\rho}<\frac{1+3w}{4}\ .\label{SE II}
\end{eqnarray}

The scaling of the energy density $\rho$, the spin density squared $S^2$ and the rate of shear squared $\sigma^2$ can be deduced respectively from the energy conservation law $\eref{En cons eq}$, the spin propagation equation $\eref{Spin cons eq}$ and the rate of shear propagation equation $\eref{Shear prop II}$ by recasting the expansion rate $\Theta$ in terms of the scale factor $R$ $\eref{scaling}$ according to,
\begin{eqnarray}
S^2=\bar{S}^2\left(\frac{R}{\bar{R}}\right)^{-6}\ ,\label{spin scale}\\
\sigma^2=\bar{\sigma}^2\left(\frac{R}{\bar{R}}\right)^{-6}\ ,\label{shear scale}\\
\rho=\bar{\rho}\left(\frac{R}{\bar{R}}\right)^{-3(1+w)}\ .\label{density scale}
\end{eqnarray}
Note that the bar corresponds to an arbitrary event (defined by a cosmic time $t=\bar{t}$), subject only to the condition $\bar{R}\neq 0$.

Furthermore, the spatial Ricci scalar $^{*}\mathcal{R}$ is the Gaussian curvature of the spatial hypersurface, which scales according to,
\begin{eqnarray}
^{*}\mathcal{R}=^{*}\bar{\mathcal{R}}\left(\frac{R}{\bar{R}}\right)^{-2}\ ,\label{curvature scale}
\end{eqnarray}  
and the cosmological constant $\Lambda$ has by definition no scale dependence,
\begin{eqnarray}
\Lambda=\bar{\Lambda}\left(\frac{R}{\bar{R}}\right)^{0}\ .\label{lambda scale}
\end{eqnarray}

Let us now assume the existence of singularities in the timelike geodesic congruence for a homogeneous and irrotational Weyssenhoff fluid. By comparing the scaling relations for the spatial Ricci scalar $\eref{curvature scale}$ and the cosmological constant $\eref{lambda scale}$ with those obtained for the spin density squared $\eref{spin scale}$ and the rate of shear squared $\eref{shear scale}$, we see that in the limit where the model tends towards a singularity (i.e. $R\rightarrow 0$), the contribution due to curvature and the cosmological constant is negligible. Hence, for a Weyssenhoff fluid with a physically reasonable equation-of-state parameter (i.e. $w<1$), the consistency $\eref{WE II}$ and singularity $\eref{SE II}$ conditions merge into a single condition according to,  
\begin{eqnarray}
\frac{S^2-\kappa^{-2}\sigma^2}{\kappa^{-1}\rho}<1\ .\label{singularity cond}
\end{eqnarray}
Moreover, we can recast this condition in terms of the scale dependence $R$. In the limit where the model tends towards a singularity, the condition $\eref{singularity cond}$ becomes,
\begin{eqnarray}
\lim_{R\rightarrow 0}\frac{\bar{S}^2-\kappa^{-2}\bar{\sigma}^2}{\kappa^{-1}\bar{\rho}}\left(\frac{R}{\bar{R}}\right)^{-3(1-w)}<1\ .\label{singularity cond limit}
\end{eqnarray}
Provided the equation-of-state parameter $w<1$, the singularity condition $\eref{singularity cond limit}$ can only hold if the rate of shear squared is larger than the spin squared (i.e. $\bar{\sigma}^2>\kappa^2\bar{S}^2$). Hence, in the opposite case, where the macroscopic spin density squared of the Weyssenhoff fluid is larger than the fluid anisotropies according to,
\begin{eqnarray}
\kappa^2S^2>\sigma^2\ ,\label{cond spin shear}
\end{eqnarray}
there will be no singularity on any scale. This is a generalisation of the result established independently for a Bianchi I metric by Kopczynski $\cite{Kopczynski:1973}$ and Stewart and Hajieck $\cite{Stewart:1973}$.

Our singularity analysis is based on the assumption that the Weyssenhoff fluid flow lines are geodesics, which implies that the macroscopic fluid (i.e. with spin averaging) has to be homogeneous. A key question is whether this still holds in presence of small inhomogeneities. According to Ellis $\cite{Ellis:2007}$, the Hawking-Penrose singularity theorems apply not only to homogeneous models but also to approximately homogeneous models with local pressure inhomogeneities. By analogy, if there is no singularity for geodesics fluid flow lines, singularities may still be averted provided the real fluid flow lines can be described as small perturbations around geodesics. 

In following sections, we will assume that the spin-shear condition $\eref{cond spin shear}$ holds, which guarantees the absence of singularities for homogeneous models.

\section{Dynamical evolution: general considerations}
\label{Dynamical evolution: general considerations}

For a homogeneous fluid, the Gaussian curvature $^{*}\mathcal{R}$ depends only on the scale factor according to,
\begin{eqnarray}
^{*}\mathcal{R}=-\frac{6k}{R^2}\ ,\label{scaling rel} 
\end{eqnarray}
where $k=\{-1,0,1\}$ is the normalised curvature parameter.

To analyse the dynamics of a homogeneous and irrotational Weyssenhoff fluid, let us first recast explicitly the Friedmann $\eref{Friedmann sing}$ and Raychaudhuri $\eref{Ricci tens eq}$ equations in terms of the physical quantities using the expression for the Gaussian curvature $\eref{scaling rel}$ according to,
\begin{eqnarray}
\left(\frac{\dot{R}}{R}\right)^2=\frac{\kappa}{3}\left[\rho-\kappa S^2+\frac{1}{\kappa}\left(\sigma^2-\frac{3k}{R^2}+\Lambda\right)\right]\ ,
\label{Large Friedmann}\\
\frac{\ddot{R}}{R}=-\frac{\kappa}{6}\left[\rho\left(1+3w\right)-4\kappa
S^2+\frac{4}{\kappa}\left(\sigma^2-\frac{1}{2}\Lambda\right)\right]\ .\label{Large Raychaudhuri}
\end{eqnarray}

We will now discuss in more details the geodesic singularities presented in $\Sref{Geodesic singularity analysis}$, drawing out more fully the geometrical and physical applications.

\subsection{Geometric interpretation of the solutions}

As outlined above, at stages of the dynamical evolution for which the scale factor $R(t)$ is small, a Weyssenhoff fluid with an equation-of-state parameter $w<1$ is dominated by the spin density and rate of shear contributions. This follows from the scaling properties of the energy $\eref{density scale}$ of the spin density $\eref{spin scale}$ and of the rate of shear $\eref{shear scale}$. Provided the spin-shear condition $\eref{cond spin shear}$ is satisfied, there can be no singularity ($R\rightarrow 0$), because the negative sign of the spin squared terms in the RHS of the Friedmann equation $\eref{Large Friedmann}$ would imply the existence of an imaginary rate of expansion, which is physically unacceptable ($\Theta\in\mathbb{R}$) as discussed before in $\Sref{Geodesic singularity analysis}$. For physical consistency, the RHS of the Friedmann equation has to be positively defined at all times,
\begin{eqnarray}
\rho-\kappa S^2+\frac{1}{\kappa}\left(\sigma^2-\frac{3k}{R^2}+\Lambda\right)\geq0\ ,
\label{Friedmann Cond}
\end{eqnarray}
which clearly excludes the presence of a singularity provided $w<1$. The physical interpretation is that $\--$ as one goes backwards in cosmic time $t$ from the present epoch $\--$ the spin contributions to the field equations dominate and produce a bounce, which we may take to occur at $t=0$, that avoids an initial singularity (i.e. $R(0)>0$). Since this model contains no initial singularity, the temporal evolution of the model, governed by the Friedmann $\eref{Large Friedmann}$ and Raychaudhuri $\eref{Large Raychaudhuri}$ equations, extends symmetrically to the negative part of the time arrow. In order to satisfy the time symmetry requirement and avoid a kink in the time evolution of the scale factor $R(t)$ at $t=0$, the expansion rate at the bounce has to vanish, $\dot{R}(0)=0$, and the temporal curvature of the scale factor $\ddot{R}(0)$ has to be finite. Thus, the scale factor $R(t)$ goes through an extremum at the bounce $R(0)=R_0$ $\footnote{Note that, throughout the paper, a zero subscript denotes the value of a quantity at the bounce (i.e. $t=0$) and not at the present epoch.}$. The energy density at the bounce, $\rho_0=\rho(0)$, is determined by the limit where the consistency requirement $\eref{Friedmann Cond}$ becomes an equality,
\begin{eqnarray}
\rho_0=\kappa S_0^2-\frac{1}{\kappa}\left(\sigma_0^2-\frac{3k}{R_0^2}+\Lambda\right)\ ,
\label{Friedmann start Cond}
\end{eqnarray}
where $S_0=S(0)$ and $\sigma_0=\sigma(0)$ denote respectively the spin energy density and the rate of shear evaluated at the bounce. Note that this particular choice for the energy density $\eref{Friedmann start Cond}$ at $t=0$ has been made in order for the expansion rate to vanish at the bounce. This can be shown explicitly by evaluating the Friedmann equation $\eref{Large Friedmann}$ at the bounce using the expression for the energy density $\eref{Friedmann start Cond}$.

Quantitative expressions or the $R(t)$-curve in various cases are derived in $\Sref{homogeneous solutions}$ below. Before doing so, however, it is worth noting that qualitatively, the general shape of the $R(t)$-curve for a Weyssenhoff fluid is closely related to the temporal curvature of the scale factor $\ddot{R}$, which is explicitly given by the Raychaudhuri equation $\eref{Large Raychaudhuri}$, and also to the range of values for $R(t)$, which is determined by the consistency condition $\eref{WE}$ on the Friedmann equation $\eref{Large Friedmann}$.

In this section, let us discuss one particular class of Weyssenhoff fluid models for which the cosmological constant $\Lambda$ is small (and positively defined),
\begin{eqnarray}
0<\Lambda\ll\rho_0\ ,
\label{lambda condition}
\end{eqnarray}
and the curvature is also small
\begin{eqnarray}
0<\frac{3}{R_0^2}\ll\rho_0\ .
\label{curvature condition}
\end{eqnarray}

The two constraints $\eref{lambda condition}$ and $\eref{curvature condition}$ on the class of models imply that the sign of the temporal curvature of the scale factor depends only on the value of the equation-of-state parameter $w$, which yields three different cases.

In the first case, where $w<-{\textstyle\frac{1}{3}}$, the RHS of the Raychaudhuri equation $\eref{Large Raychaudhuri}$ implies that the temporal curvature of the scale factor is positively defined at all times,
\begin{eqnarray}
\ddot{R}(t)>0\ \ \ \ \mathrm{for} \ \ \ t\in(-\infty,\infty)\ .\label{curvature 1}
\end{eqnarray}
The positive sign of $\ddot{R}$ implies that the scale factor is minimal at the bounce and the model is perpetually inflating (for $t>0$).  

In the second case, where $w>1$, by comparing the consistency requirement $\eref{Friedmann Cond}$ with the Raychaudhuri equation $\eref{Large Raychaudhuri}$, the temporal curvature of the scale factor is found to be negatively defined at all times,
\begin{eqnarray}
\ddot{R}(t)<0\ \ \ \ \mathrm{for} \ \ \ t\in(-\infty,\infty)\ .\label{curvature 2}
\end{eqnarray} 
Note that for a model with an equation-of-state parameter $w>1$, we reach the same conclusion as for a fluid with an equation of state parameter $w<1$, which is that the model has a time-symmetric evolution and bounces at $t=0$. The negative sign of $\ddot{R}$ implies that the scale factor is maximal at the bounce and is deflating (for $t>0$) until it eventually collapses.

In the third case, where $-{\textstyle\frac{1}{3}}<w<1$, the symmetric time evolution of the scale factor can be split into five parts. Firstly, for a small cosmic time, i.e. $|t|<|t_f|$ $\--$ where the value of $t_f$ depends on the scale parameter $w$ $\--$ the sign of the temporal curvature of the scale factor is positive. This corresponds to the spin dominated phase. Secondly, for a specific cosmic time, i.e. $|t|=|t_f|$, the temporal curvature of the scale factor vanishes as the time evolution of the scale factor reaches an inflection point. Then, for a larger cosmic time, i.e. $|t_f|<|t|<|t_a|$, the temporal curvature of the scale factor has the opposite sign until it reaches the second inflection point $|t|=|t_a|$. This corresponds to the matter dominated phase. Finally, for large cosmic time, i.e. $|t|>|t_a|$, the sign of the temporal curvature of the scale factor becomes positive again. This corresponds to the cosmological constant dominated phase. The behaviour of $\ddot{R}(t)$ in terms of cosmic time $t$ is summarised as follows,
\begin{eqnarray}
\ddot{R}(t)>0\ \ \ \ \mathrm{for} \ \ \ t\in(-t_f,t_f)\ , \label{a}\\
\ddot{R}(t)=0\ \ \ \ \mathrm{for} \ \ \ t\in\{-t_f,t_f\}\ ,\label{b}\\
\ddot{R}(t)<0\ \ \ \ \mathrm{for} \ \ \ t\in(-t_a,-t_f)\cup(t_f,t_a)\ ,\label{c}\\
\ddot{R}(t)=0\ \ \ \ \mathrm{for} \ \ \ t\in\{-t_a,t_a\}\ ,\label{d}\\
\ddot{R}(t)>0\ \ \ \ \mathrm{for} \ \ \ t\in(-\infty,-t_a)\cup(t_a,\infty)
\ .\label{e}
\end{eqnarray}

In the first and second cases, the results obtained for the symmetric time evolution of the scale factor are interesting mathematical solutions, but they are inconsistent with current cosmological observations. In order to satisfy the current cosmological data, the positively defined time evolution of the model has to inflate, i.e. $\ddot{R}(t)>0$, at early time ($t<t_f$), and produce a sufficient amount of inflation. At later time ($t>t_f$), the energy density of the fluid dominates the dynamics and acts like a brake on the expansion $\ddot{R}(t)<0$.

During the spin-dominated phase, the contribution due to the cosmological constant can be safely neglected $\eref{lambda condition}$ and the positive temporal curvature of the scale factor $\eref{a}$ leads to an inflation phase. The inflatability condition, $\ddot{R}(t)>0$, may be deduced from the Raychaudhuri equation $\eref{Large Raychaudhuri}$ according to,
\begin{eqnarray}
\rho(1+3w)-4\kappa S^2+4\kappa^{-1}\sigma^2<0\ .
\label{Raychaudhuri Cond}
\end{eqnarray}
This inflation phase ends when this inequality is no longer satisfied, which corresponds to the inflection point of the temporal evolution of the scale factor, i.e. $t=t_f$. Hence, at the end of inflation the density is given by,
\begin{eqnarray}
\rho_f=\frac{4\kappa}{(1+3w)}\left(S_f^2-\kappa^{-2}\sigma_f^2\right)\
. \label{Raychaudhuri end Cond}
\end{eqnarray}

The temporal evolution of this model for a positively defined time is characterised by a maximal physical energy density $\rho=\rho_0$ coinciding with the start of an inflation phase ending when the energy density reaches the density threshold $\rho=\rho_f$. At the end of inflation, the model enters a matter dominated phase. During this stage, the Weyssenhoff fluid model reduces asymptotically to the cosmological solution obtained for a perfect fluid in GR in the limit where the cosmic time is sufficiently large $t\gg t_f$, which eventually leads to a cosmological constant dominated phase for $t\gg t_a>t_f$.

\subsection{Amount of inflation}

The amount of inflation is measured by the number $N$ of e-folds, which is determined using the scaling of the energy density $\eref{density scale}$, the initial $\eref{Friedmann start Cond}$ and final $\eref{Raychaudhuri end Cond}$ energy densities, and found to be,
\begin{eqnarray}
N\equiv
\mathrm{ln}\frac{R_f}{R_0}=-\frac{1}{3(1+w)}\mathrm{ln}\left[\frac{4}{1+3w}
\left(\frac{\kappa^{2}S_f^2-\sigma_f^2}
{\kappa^{2}S_0^2-\sigma_0^2}\right)\right]\ .
\label{E-fold}
\end{eqnarray}

Using the scaling relations obtained for the spin density squared $\eref{spin scale}$ and the rate of shear squared $\eref{shear scale}$, the initial and final values of these quantities are found to be related by the number of e-folds according to,
\begin{eqnarray}
S_0^2=S_f^2\left(\frac{R_0}{R_f}\right)^{-6}=S_f^2e^{6N}\ ,\label{scaling efold spin}\\
\sigma_0^2=\sigma_f^2\left(\frac{R_0}{R_f}\right)^{-6}=\sigma_f^2e^{6N}\ .\label{scaling efold shear}
\end{eqnarray} 

\begin{figure}[htbp]
\begin{center}
\includegraphics[width=11cm,height=9cm]{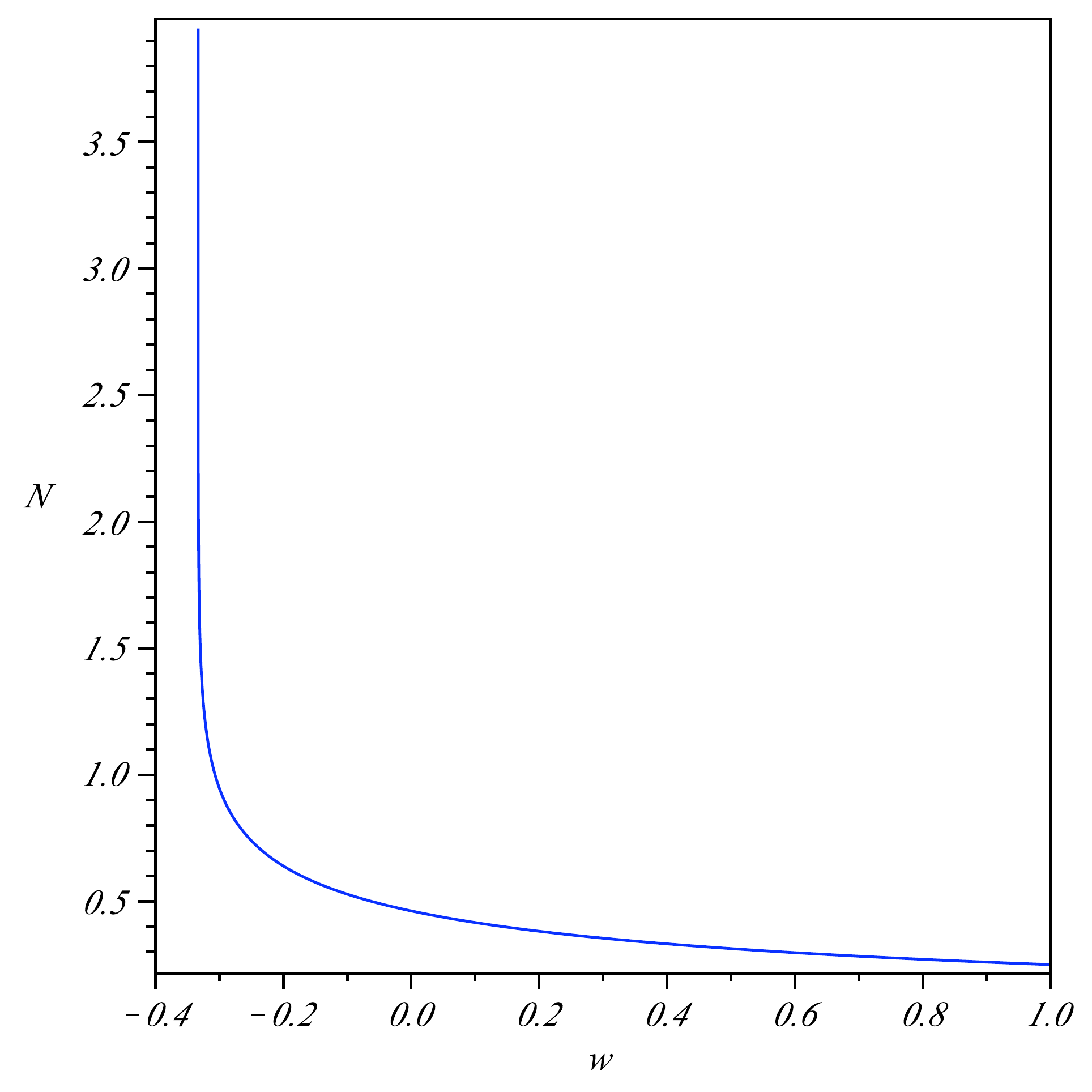}
\caption{Number of e-folds $N$ in terms of the equation-of-state parameter $w$. $N(w)$ has a vertical asymptote at $w=-{\textstyle\frac{1}{3}}$.\label{fig e-folds}}
\end{center}
\end{figure}

By recasting the initial values of the spin density squared and rate of shear squared in terms of their final values according to $\eref{scaling efold spin}$ and $\eref{scaling efold shear}$ respectively, the expression for the number of e-folds $\eref{E-fold}$ reduces to an elegant expression, 
\begin{eqnarray}
N=\frac{1}{3(1-w)}\mathrm{ln}\left(\frac{4}{1+3w}\right)\ ,
\label{e-fold}
\end{eqnarray}
and is shown in $\Fref{fig e-folds}$. It worth mentioning that the amount of inflation is independent of the rate of shear or the spin density of the fluid.
Let us mention that Bianchi models based on a Weyssenhoff fluid have been studied previously by Lu and Cheng $\cite{Lu:1995}$. However, the authors did not try to estimate the amount of inflation in their analysis.

The only way to have achieve a substantial number of e-folds is by requiring an equation of state of the form
\begin{eqnarray}
w=-\frac{1}{3}+\epsilon \ \ \ \ \ \mathrm{where}\ \ \ \ \
0<\epsilon\ll 1\ ,\label{State tunning}
\end{eqnarray}
which corresponds to no standard fluid and has therefore no acceptable physical basis. This conclusion has already been reached by Gasperini $\cite{Gasperini:1986}$ in the isotropic case. We have showed that the same result still holds in the anisotropic case. 

It is interesting to note that a cosmic string fluid has an equation-of-state parameter $w=-{\textstyle\frac{1}{3}}$. A hybrid Weyssenhoff fluid made for example of fermionic matter cosmic strings $\cite{Ringeval:2000}$ and matter fields $\--$ where the cosmic strings contribution dominates the dynamics at the era of interest $\--$ has an equation-of-state parameter of the form $\eref{State tunning}$ where the value of the fine tuning parameter $\epsilon$ depends crucially on the ratio between the cosmic string and the matter fields densities. Although such a fluid is a candidate to obtain an inflation phase at an early positively defined time (i.e. just after the bounce), it does not reduce to the cosmological standard model at later times when the spin contribution can be safely neglected. This is due to the fact that the density of the cosmic strings contribution $\rho_{\mathrm{st}}$ scales as $\rho_{\mathrm{st}}\propto R^{-2}$. Hence, if the cosmic strings contribution dominates the behaviour of the cosmic fluid for an early positively defined time, it will do so at all times. 

However, this problem may potentially be overcome by assuming that, at early times, the cosmic strings decay into the matter fields of the standard model leading to a reheating phase. It would be worth further investigating this possibility.  

The fine tuning parameter $\epsilon$ has a magnitude that is related to the number of e-folds according to,
\begin{eqnarray}
\epsilon\sim e^{-4N}\ .\label{State tunned}
\end{eqnarray}
To obtain, for example, an inflationary phase with $N=50-70$ e-folds $\--$ which is a characteristic range of values for current parameter estimations $\--$ the equation of state has to be very fine tuned such that $\epsilon=10^{-87}-10^{-122}$. It is worth noting that this is a similar order of magnitude to the factor $10^{-120}$ relating the ratio of the cosmological constant predicted by summing the zero point energy of the Standard Model fields up to the Planck cutoff to that inferred from cosmological observations, although this is almost certainly just a numerical coincidence.

\section{Quantitative dynamical evolution of spatially-curved models}
\label{Quantitative dynamical evolution of spatially-curved models}

Our general approach allows one to investigate models with non-zero spatial curvature and a cosmological constant. In general, it is not possible to find analytical solutions for the time evolution of the scale factor. However, the behaviour of the solutions can be analysed by integrating the dynamical equations numerically. The analysis and plots of the time evolution of the scale factor in spatially-curved models are presented below.

\subsection{Solutions in presence of a cosmological constant}
\label{cosmological constant}

The dynamics of a homogeneous and anisotropic Weyssenhoff fluid in a spatially-curved model in presence of a cosmological constant relies on the Fridemann $\eref{Large Friedmann}$ and Raychaudhuri $\eref{Large Raychaudhuri}$ equations. Using the scaling relation obtained for the energy density $\eref{density scale}$, for the spin density $\eref{spin scale}$, and for rate of shear $\eref{shear scale}$, the Friedmann $\eref{Large Friedmann}$ and Raychaudhuri $\eref{Large Raychaudhuri}$ equations can be recast respectively as,
\begin{eqnarray}
\fl\left(\frac{\dot{R}}{R}\right)^2=\frac{\kappa}{3}\rho_0\left({\frac{R}{R_0}}\right)^{-3(1+w)}-\frac{\kappa^2}{3}\left(S_0^2-\kappa^{-2}\sigma_0^2\right)\left({\frac{R}{R_0}}\right)^{-6}-\frac{k}{R_0^2}\left({\frac{R}{R_0}}\right)^{-2}+\frac{\Lambda}{3}\ ,\label{Friedmann initial gene}\\
\fl\;\phantom{\Big(}\frac{\ddot{R}}{R}\phantom{\Big)^2}=-\frac{\kappa}{6}\rho_0\left(1+3w\right)\left(\frac{R}{R_0}\right)^{-3(1+w)}+\frac{2}{3}\kappa^2\left(S_0^2-\kappa^{-2}\sigma_0^2\right)\left(\frac{R}{R_0}\right)^{-6}+\frac{\Lambda}{3}\ ,\label{Raychaudhuri initial gene}
\end{eqnarray}
where for $t=0$, $R_0$ is the scale factor, $\rho_0$ the energy density, $S_0$ the spin density and $\sigma_0$ the rate of shear.

For convenience, we introduce six dimensionless parameters defined as, 
\begin{eqnarray}
r\equiv\frac{R}{R_0}\ ,\\
\tau\equiv\sqrt{\frac{\kappa\rho_0}{3}}t\ ,\\
\delta^2\equiv\frac{\sigma_0^2}{\kappa\rho_0}\ ,\label{delta}\\
s^2\equiv\frac{\kappa S_0^2}{\rho_0}\ ,\label{s^2}\\
\alpha\equiv\frac{3k}{\kappa\rho_0R_0^2}\ ,\label{alpha}\\
\lambda\equiv\frac{\Lambda}{\kappa\rho_0}\ ,\label{lambda}
\end{eqnarray}
which are the scale factor parameter $r$, the cosmic time parameter $\tau$, the rate of shear squared parameter $\delta^2$ and the spin density squared parameter $s^2$, the curvature parameter $\alpha$, the cosmological constant parameter $\lambda$. Note that $r$ and $\tau$ depend on $t$, whereas $\delta^2$, $s^2$, $\alpha$ and $\lambda$ are constant, defined in terms of quantities at the bounce $t=0$.

The consistency condition at the bounce $\eref{Friedmann start Cond}$ can be recast in terms of dimensionless parameters as,
\begin{eqnarray}
s^2-\delta^2=1-\alpha+\lambda\ .
\label{Derivability cond lambda}
\end{eqnarray}
Using $\eref{Derivability cond lambda}$, the Friedmann $\eref{Friedmann initial gene}$ and Raychaudhuri $\eref{Raychaudhuri initial gene}$ equations can also be recast respectively in terms of the dimensionless parameters according to,
\begin{eqnarray}
{r^{\prime}}^2=\frac{1}{r^4}\left(r^{3(1-w)}-\alpha\left(r^4-1\right)-1+\lambda\left(r^6-1\right)\right)\ ,\label{Friedmann lambda}\\
r^{\prime\prime}=-\frac{2}{r^5}\left(\frac{1+3w}{4}r^{3(1-w)}+\alpha-1-\lambda\left(\frac{r^6}{2}+1\right)\right)\ ,\label{Raychaudhuri lambda}
\end{eqnarray}
where a prime denotes a derivative with respect to the rescaled cosmic time parameter $\tau$. It worth emphasizing that the dynamics of a homogeneous Weyssenhoff fluid does not depend explicitly on the rate of shear parameter squared $\delta^2$. This is due to the fact that the rate of shear $\eref{shear scale}$ scales like the spin density $\eref{spin scale}$, and follows explicitly from the fact that the $\delta^2-$terms cancel after substituting the consistency condition at the bounce $\eref{Derivability cond lambda}$ into the dynamical equations $\eref{Friedmann lambda}$ and $\eref{Raychaudhuri lambda}$. However, the consistency condition implies that the value of the physical quantities at the bounce still depends on the corresponding value of rate of shear. 

The physical interpretation of these equations is well known. The Friedmann equation corresponds to the conservation law of energy whereas the Raychaudhuri equation represents the equation of motion.

The Friedmann equation $\eref{Friedmann lambda}$ can be recast as follows,
\begin{equation}
\frac{1}{2}{r^{\prime}}^2+U_{\mathrm{eff}}(r)=-\frac{\alpha}{2}\ ,
\end{equation}
where the effective potential is given by
\begin{equation}
U_{\mathrm{eff}}(r)=-\frac{1}{2r^4}\left(r^{3(1-w)}+\alpha-1+\lambda\left(r^6-1\right)\right)\ .\label{eff pot}
\end{equation}

The parameters present in the Friedmann and Raychaudhuri equations are respectively,
\begin{itemize}
\item $w$: relativistic pressure (SR: continuous parameter),
\item $\alpha$: curvature (GR: continuous parameter),
\item $-1$: spin (EC: discrete parameter),
\item $\lambda$: cosmological constant.
\end{itemize}

From the expression for the effective potential $\eref{eff pot}$, we see that the spin contribution has a positive sign, which means that it behaves like a potential barrier. In other words, the spin-spin interaction leads to repulsive centrifugal forces opposing the attractive effect of gravity, thus preventing collapse. Note that this is also the case for a positive cosmological constant.   

In the absence of relativistic pressure (i.e. $w=0$), curvature (i.e. $\alpha=0$), spin (i.e. the $-1$ factor vanishes), and cosmological constant (i.e. $\lambda=0$) the Friedmann and the Raychaudhuri equations reduce respectively to the energy conservation law for a particle in a gravitational field with a vanishing total energy ($E_{\mathrm{tot}}=0$), and Newton's second law of motion.

The mathematical solutions for the time evolution of scale factor parameter depend on the whole real range of the parameters (i.e. $w, \alpha, \lambda \in \mathbb{R}$). But for physical consistency, we have to restrict the value of these parameters. Firstly, the Weyssenhoff fluid cannot violate causality (i.e. $c_s<c$), which sets an upper bound on the equation-of-state parameter $w$, 
\begin{equation}
w<1\ .\label{upper equ state}
\end{equation}
Secondly, the spin-shear condition $\eref{cond spin shear}$ and the consistency condition at the bounce $\eref{Derivability cond lambda}$ restrict the range of the cosmological and curvature parameters according to,
\begin{equation}
\lambda>\alpha-1\ .\label{consistency lambda}
\end{equation}

In general, it is not possible to find analytic solutions for the Friedmann $\eref{Friedmann lambda}$ and Raychaudhuri $\eref{Raychaudhuri lambda}$ equations. However, it is possible to deduce the behaviour of the solutions by studying the asymptotic behaviour of the expansion rate parameter $r^{\prime}$ and its derivative $r^{\prime\prime}$.

In the limit where $r\rightarrow 1$, the temporal curvature of the scale factor behaves like,
\begin{equation}
\lim_{r \to 1}r^{\prime\prime}=-3\left(\textstyle{\frac{2}{3}}\alpha +\textstyle{\frac{1}{2}}\left(w-1\right)-\lambda\right)\ ,\label{i}
\end{equation}
and the expansion rate parameter $r^{\prime}$ has to vanish, 
\begin{equation}
\lim_{r \to 1}r^{\prime}=0\ ,
\end{equation}
to satisfy the consistency condition at the bounce $\eref{Derivability cond lambda}$. Hence, we find three types of solutions which depend on the respective value of the parameters:
\begin{enumerate}
	\item $\lim_{r \to 1}r^{\prime\prime}>0$, which implies that the solution $r(\tau)$ is found within the range
	\begin{equation}
	1\leq r<\infty\ ,
	\end{equation}
    provided the parameters $w$ and $\alpha$ satisfy
	\begin{equation}
	\lambda>\textstyle{\frac{2}{3}}\alpha +\textstyle{\frac{1}{2}}\left(w-1\right)\ .\label{ii}
	\end{equation}
	\item $\lim_{r \to 1}r^{\prime\prime}=0$, which implies that the solution $r(\tau)$ is static
	\begin{equation}
	r=1\ ,
	\end{equation}
	when the parameters $w$ and $\alpha$ satisfy
	\begin{equation}
	\lambda=\textstyle{\frac{2}{3}}\alpha +\textstyle{\frac{1}{2}}\left(w-1\right)\ .\label{iii}
	\end{equation}
	\item $\lim_{r \to 1}r^{\prime\prime}<0$, which implies that the solution $r(\tau)$ is found within the range
	\begin{equation}
	0\leq r\leq 1\ ,
	\end{equation}
	provided the parameters $w$ and $\alpha$ satisfy
	\begin{equation}
	\lambda<\textstyle{\frac{2}{3}}\alpha +\textstyle{\frac{1}{2}}\left(w-1\right)\ .\label{iiii}
	\end{equation}
	Moreover, the limit, $\lim_{r \to 0}{r^{\prime}}^{2}=-\infty< 0$, clearly does not exist. Hence, the solutions always satisfy $r>0$, which means that there cannot be any singularity. Thus, for a negative temporal curvature $r^{\prime\prime}<0$, the scale factor $r$ reaches a minimum value $r^*$ found within the range $0<r^*<1$.
\end{enumerate}

\begin{figure}[htp]
	\begin{minipage}[b]{0.5\linewidth}
		\centering
		\includegraphics[scale=0.35]{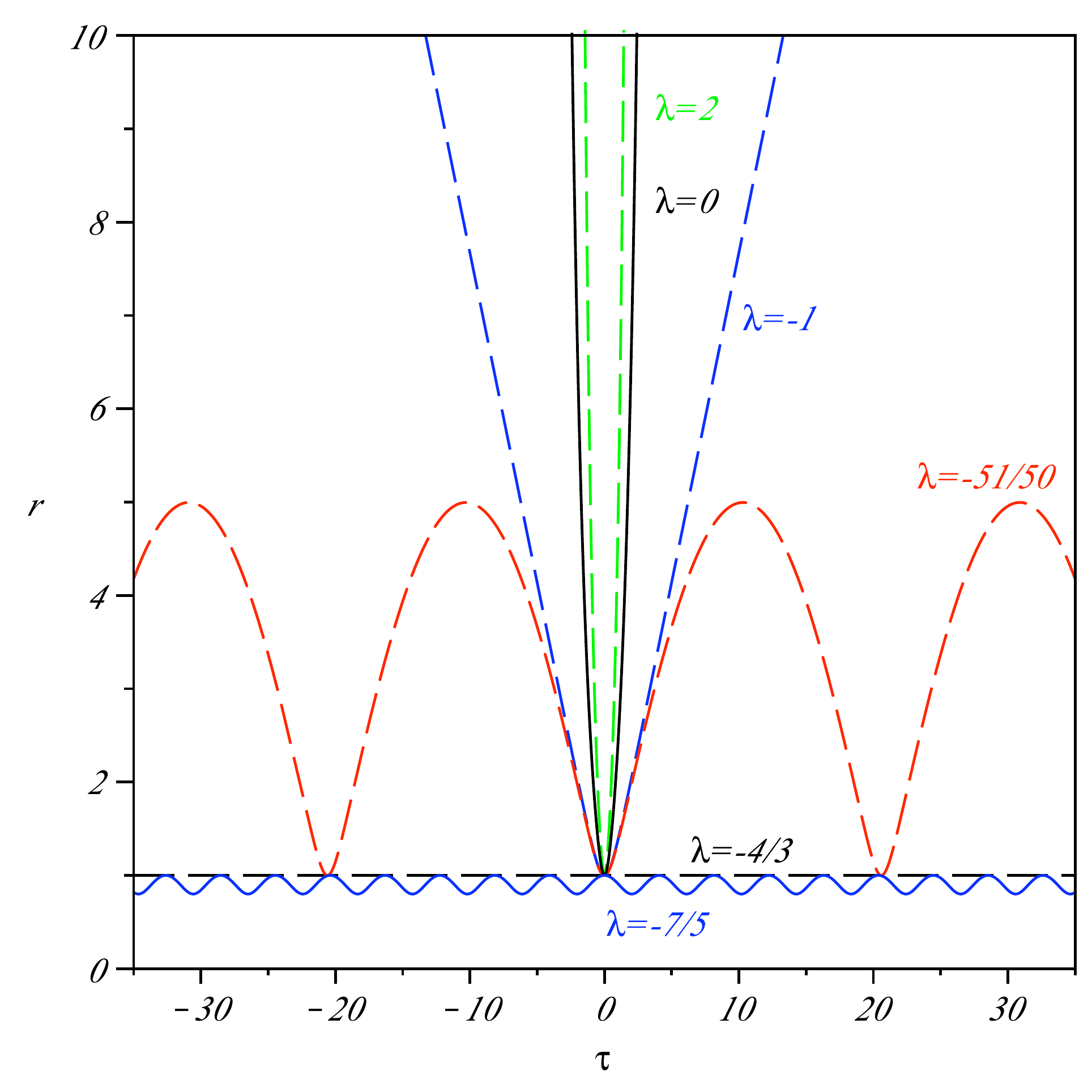}
		\caption{$\left(w=-1, \alpha={-\textstyle\frac{1}{2}}\right)$: $r(\tau)$ curves for $\qquad\qquad\qquad$ $\lambda=\{-{\textstyle\frac{7}{5}}, -{\textstyle\frac{4}{3}}, -{\textstyle\frac{51}{50}}, -1, 0, 2\}$}
		\label{w_-1_a_-1_2}
	\end{minipage}
	\hspace{0.5cm}
	\begin{minipage}[b]{0.5\linewidth}
		\centering
		\includegraphics[scale=0.35]{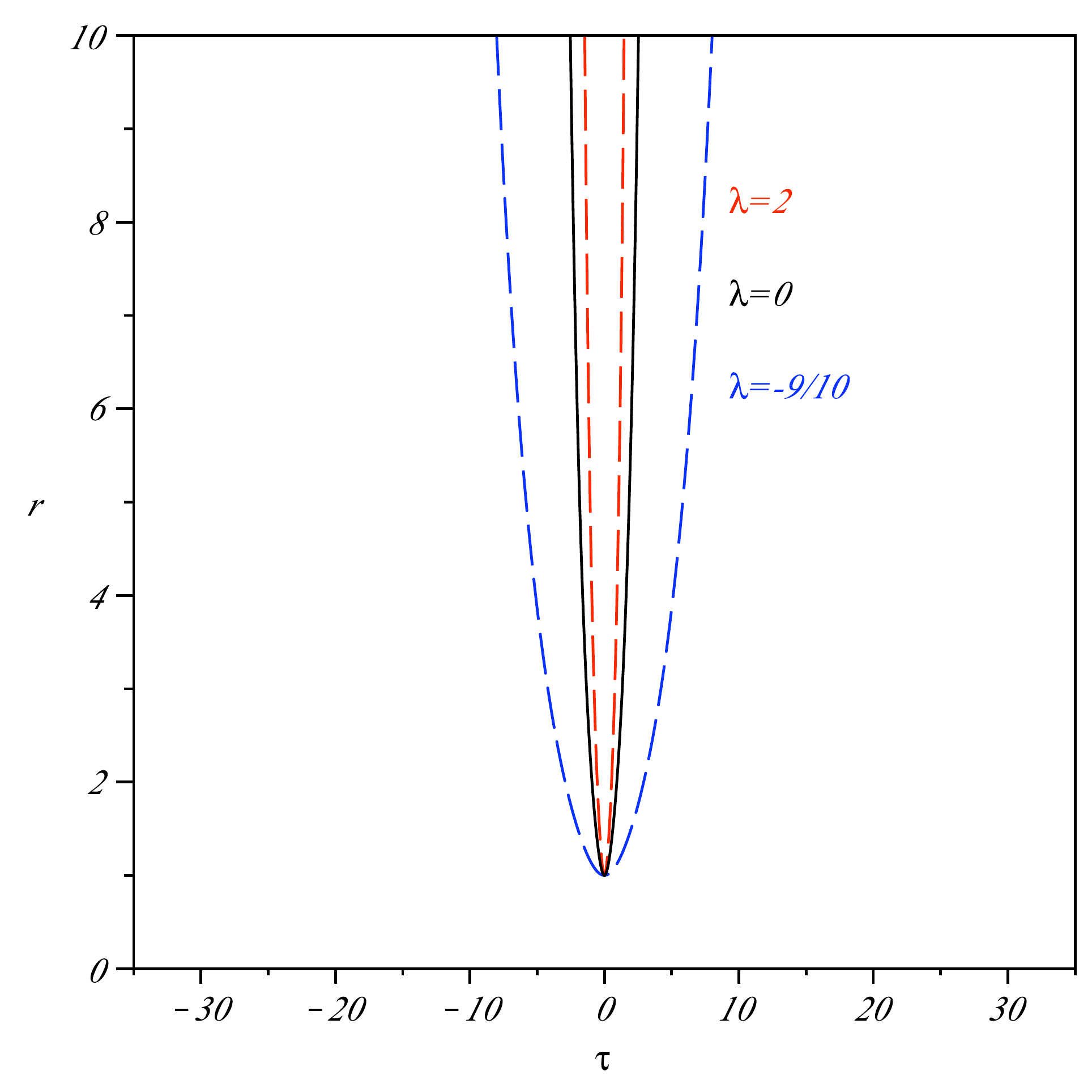}
		\caption{$\left(w=-1, \alpha=0\right)$: $r(\tau)$ curves for $\lambda=\{{-\textstyle\frac{9}{10}}, 0, 2\}$}
		\label{w_-1_a_0}
	\vspace{0.5cm}
	\end{minipage}
\end{figure}

\begin{figure}
	\begin{minipage}[b]{0.5\linewidth}
		\centering
		\includegraphics[scale=0.35]{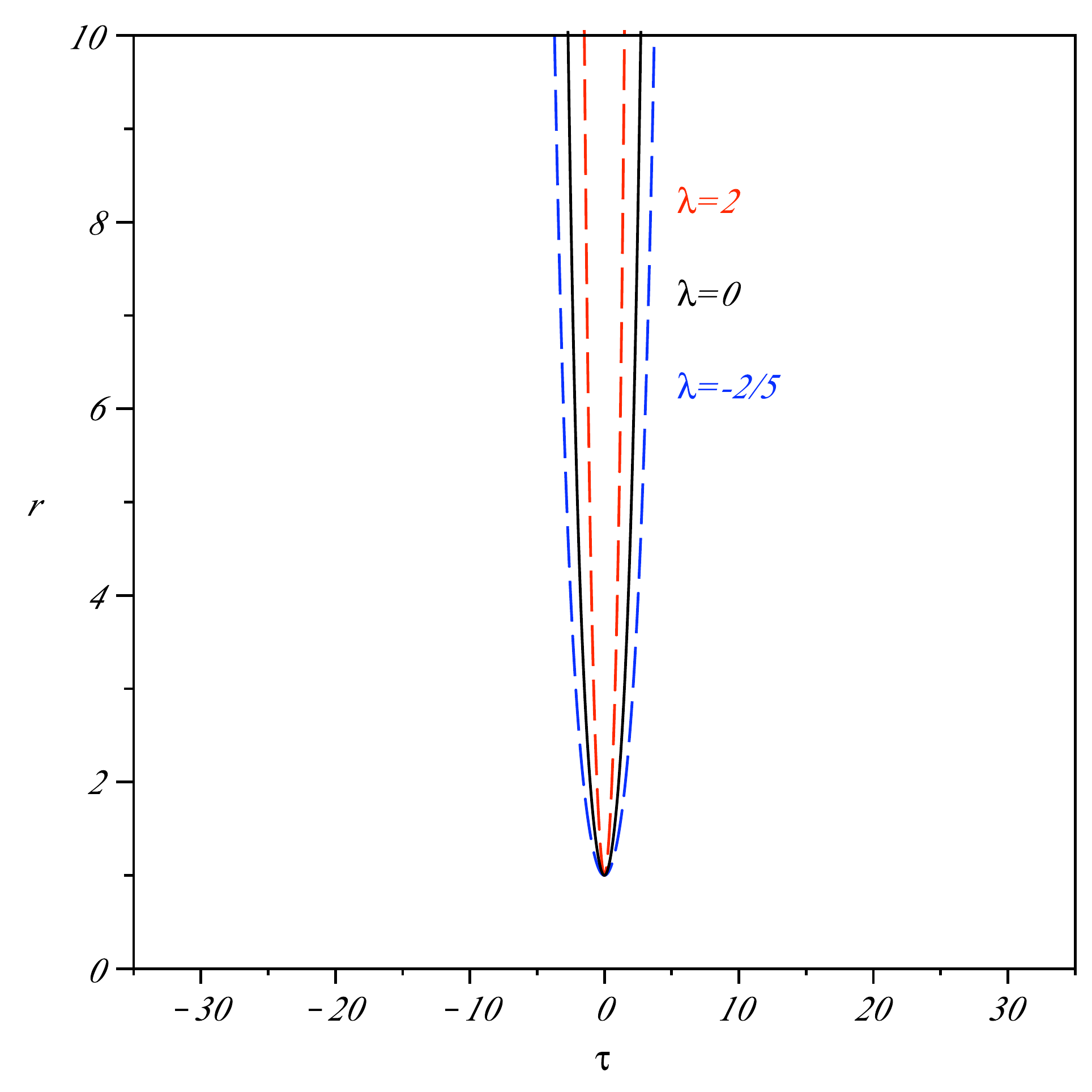}
		\caption{$\left(w=-1, \alpha={\textstyle\frac{1}{2}}\right)$: $\quad$ $r(\tau)$ curves for $\lambda=\{{-\textstyle\frac{2}{5}}, 0, 2\}$}
		\label{w_-1_a_1_2}
	\vspace{2 cm}
	\end{minipage}
	\hspace{0.5cm}
	\begin{minipage}[b]{0.5\linewidth}
		\centering
		\includegraphics[scale=0.35]{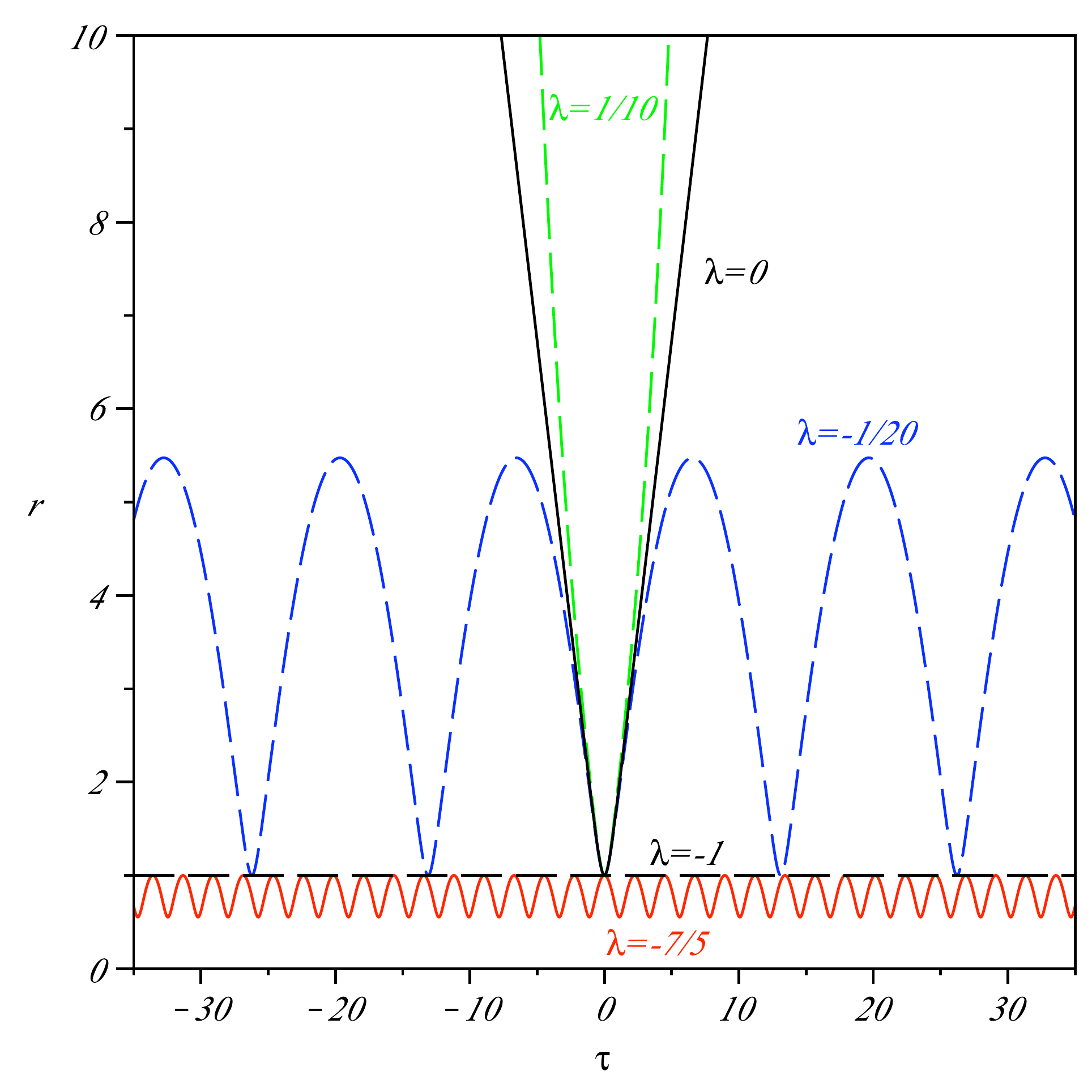}
		\caption{$\left(w=-{\textstyle\frac{1}{3}}, \alpha={-\textstyle\frac{1}{2}}\right)$: $r(\tau)$ curves for $\qquad\qquad\qquad$ $\lambda=\{-{\textstyle\frac{7}{5}}, -1, -{\textstyle\frac{1}{20}}, 0, {\textstyle\frac{1}{10}}\}$}
		\label{w_-1_3_a_-1_2}
	\vspace{1.5cm}
	\end{minipage}
	\begin{minipage}[b]{0.5\linewidth}
		\centering
		\includegraphics[scale=0.35]{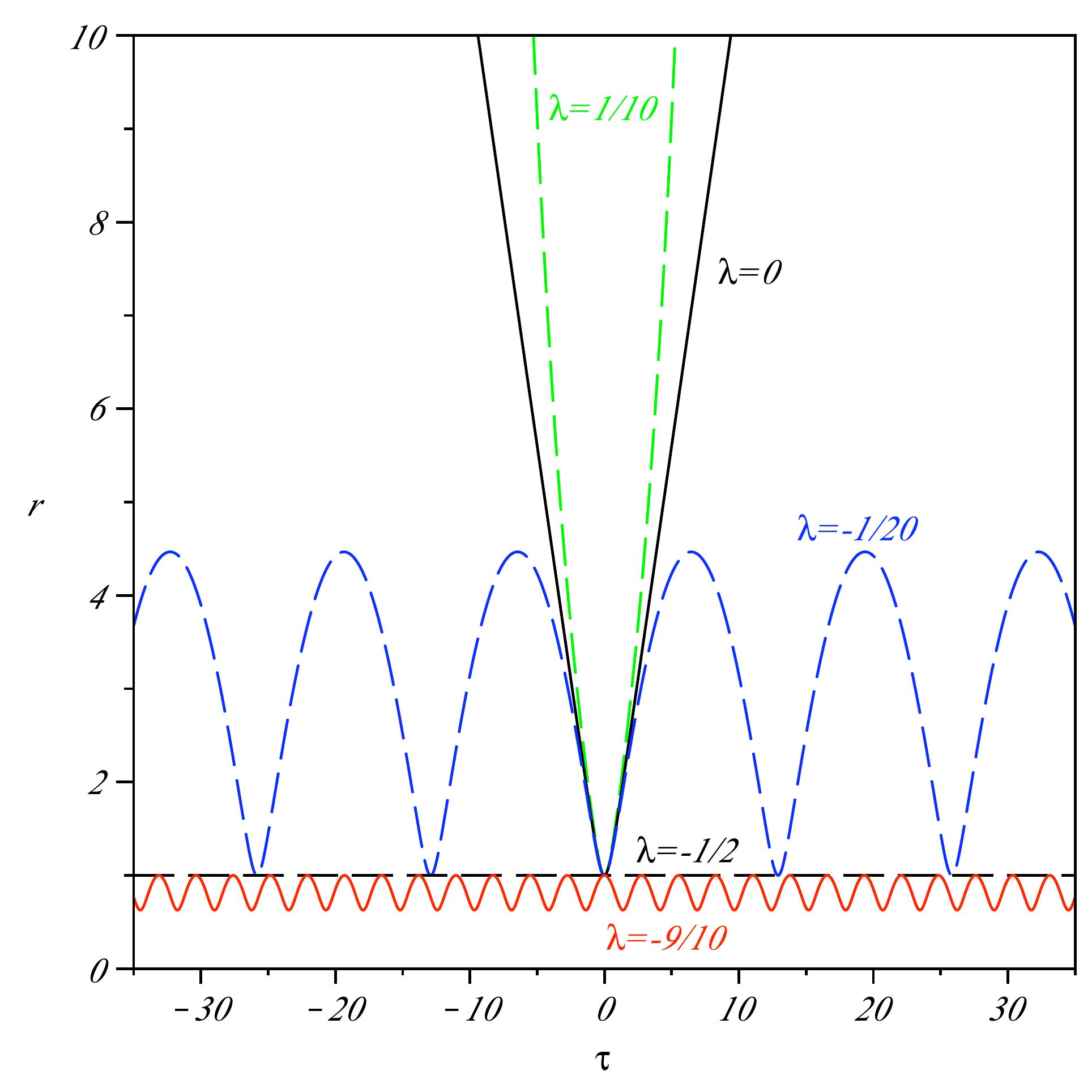}
		\hspace{-1cm}\caption{$\left(w=-{\textstyle\frac{1}{3}}, \alpha=0\right)$: $\quad$ $r(\tau)$ curves for $\qquad\qquad\qquad$ $\lambda=\{{-\textstyle\frac{9}{10}}, -{\textstyle\frac{1}{2}}, -{\textstyle\frac{1}{20}}, 0, {\textstyle\frac{1}{10}}\}$}
		\label{w_-1_3_a_0}
	\vspace{1.5cm}
	\end{minipage}
	\hspace{0.5cm}
	\begin{minipage}[b]{0.5\linewidth}
		\centering
		\includegraphics[scale=0.35]{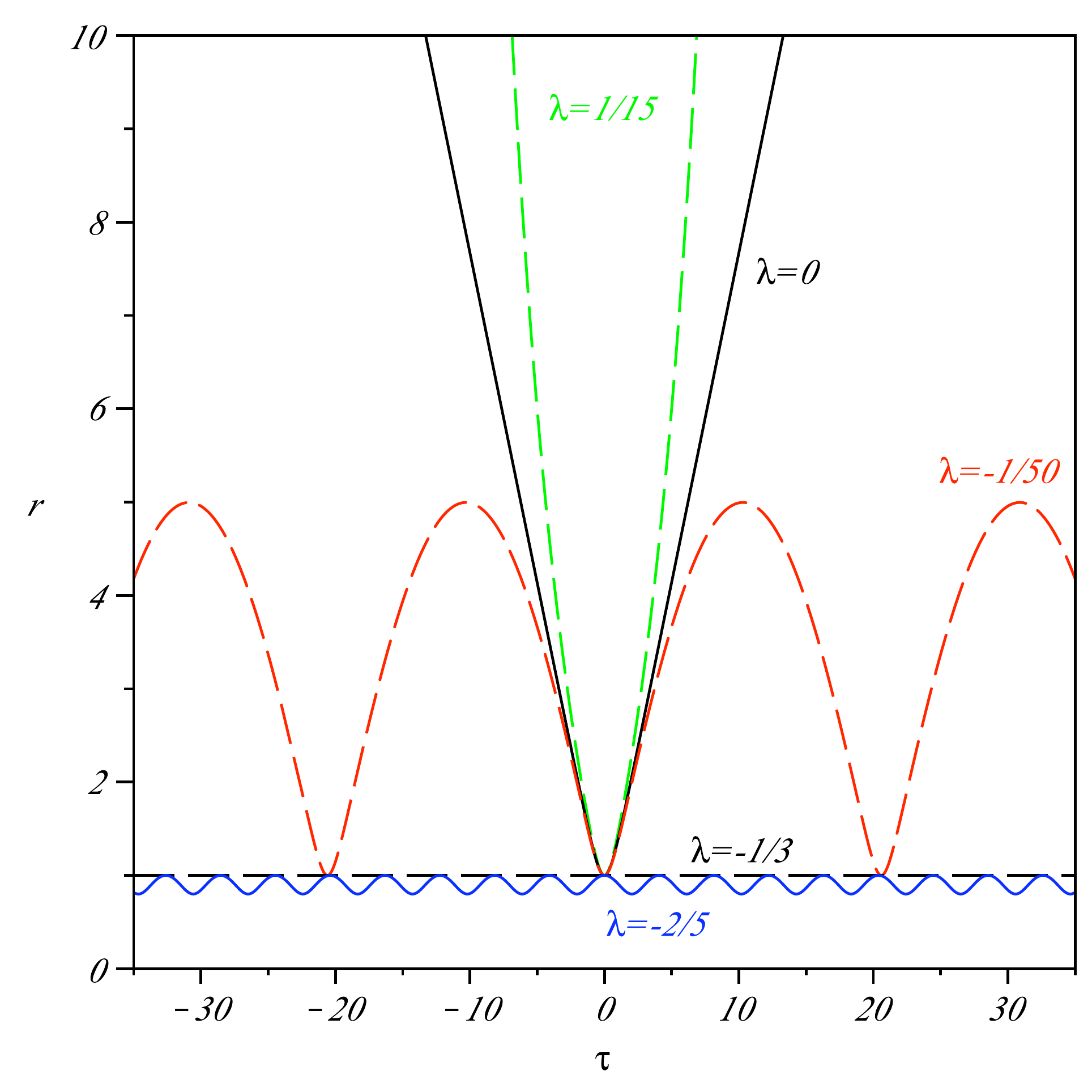}
		\caption{$\left(w=-{\textstyle\frac{1}{3}}, \alpha={\textstyle\frac{1}{2}}\right)$: $\quad$ $r(\tau)$ curves for $\qquad\qquad\qquad$ $\lambda=\{{-\textstyle\frac{2}{5}}, {-\textstyle\frac{1}{3}}, {-\textstyle\frac{1}{50}}, 0, {\textstyle\frac{1}{15}}\}$}
		\label{w_-1_3_a_1_2}
	\vspace{1.5cm}
	\end{minipage}
\end{figure}

\begin{figure}
	\begin{minipage}[b]{0.5\linewidth}
		\centering
		\includegraphics[scale=0.35]{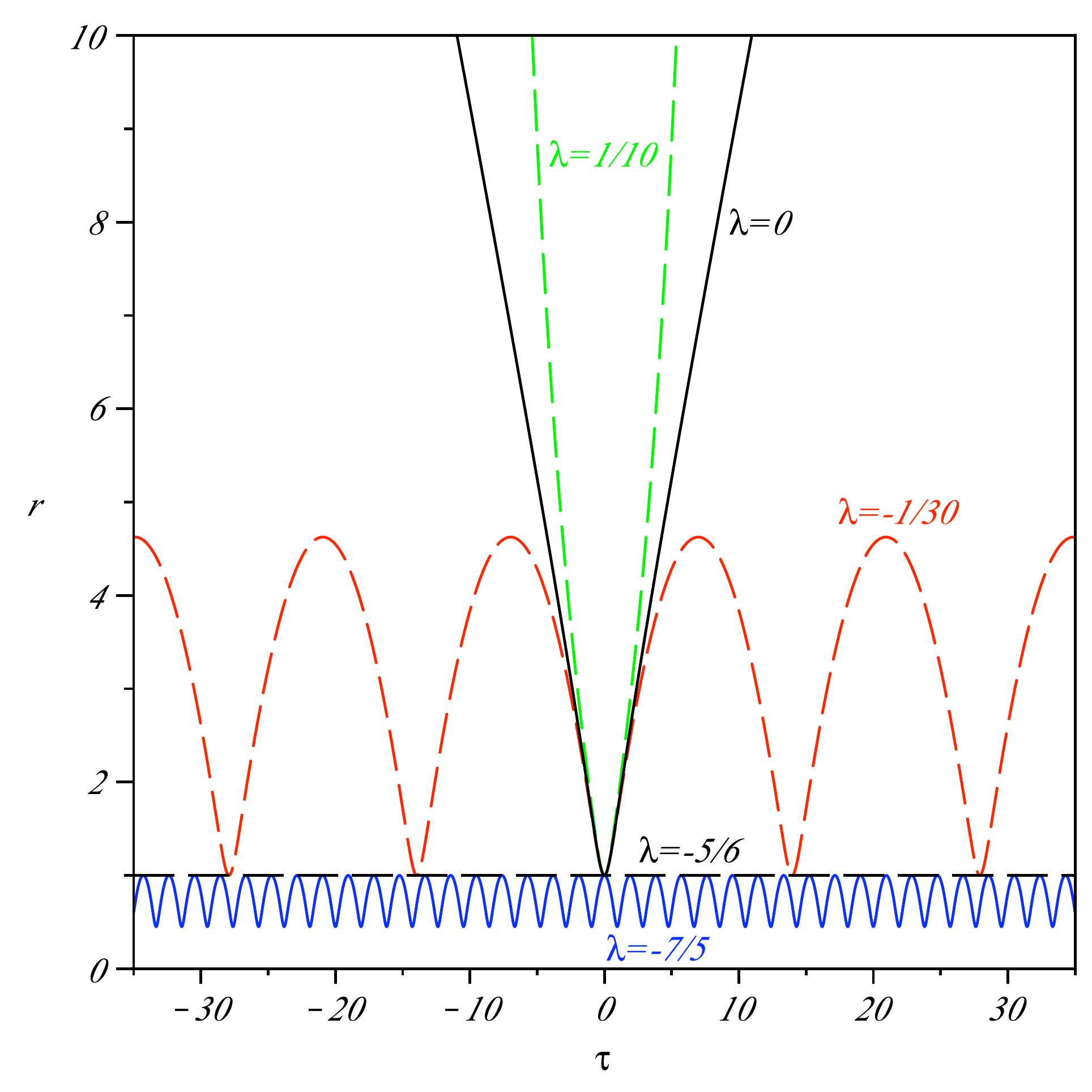}
		\caption{$\left(w=0, \alpha=-{\textstyle\frac{1}{2}}\right)$: $\quad$ $r(\tau)$ curves for $\qquad\qquad\qquad$ $\lambda=\{-{\textstyle\frac{7}{5}}, -{\textstyle\frac{5}{6}}, -{\textstyle\frac{1}{30}}, 0, {\textstyle\frac{1}{10}}\}$}
		\label{w_0_a_-1_2}
	\vspace{1.5cm}
	\end{minipage}
	\hspace{0.5cm}
	\begin{minipage}[b]{0.5\linewidth}
		\centering
		\includegraphics[scale=0.35]{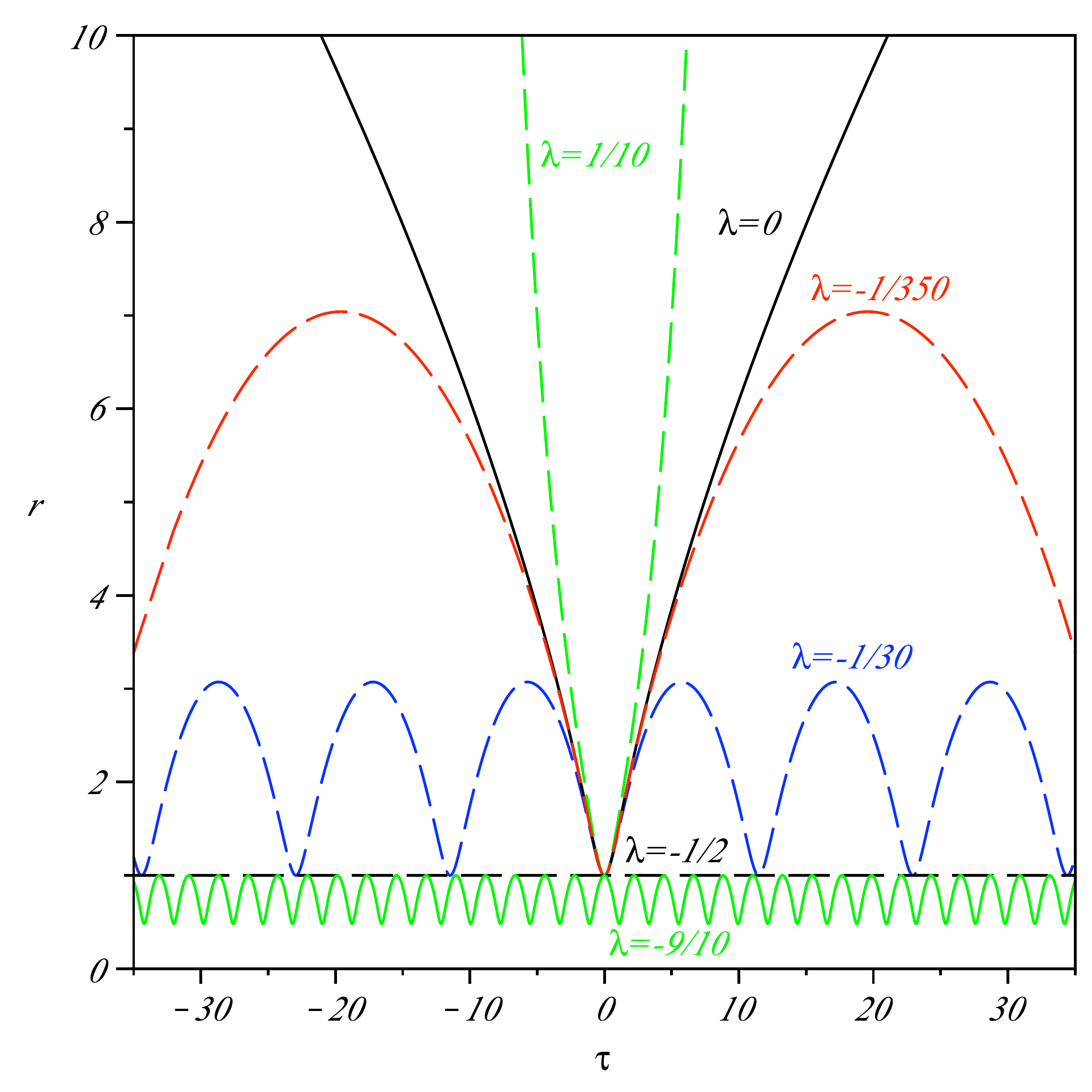}
		\caption{$\left(w=0, \alpha=0\right)$: $\quad\quad$ $r(\tau)$ curves for $\qquad\qquad\qquad$ $\lambda=\{-{\textstyle\frac{9}{10}}, -{\textstyle\frac{1}{2}}, -{\textstyle\frac{1}{30}}, -{\textstyle\frac{1}{350}}, 0, {\textstyle\frac{1}{10}}\}$}
		\label{w_0_a_0}
	\vspace{1.5cm}
	\end{minipage}
	\begin{minipage}[b]{0.5\linewidth}
		\centering
		\includegraphics[scale=0.35]{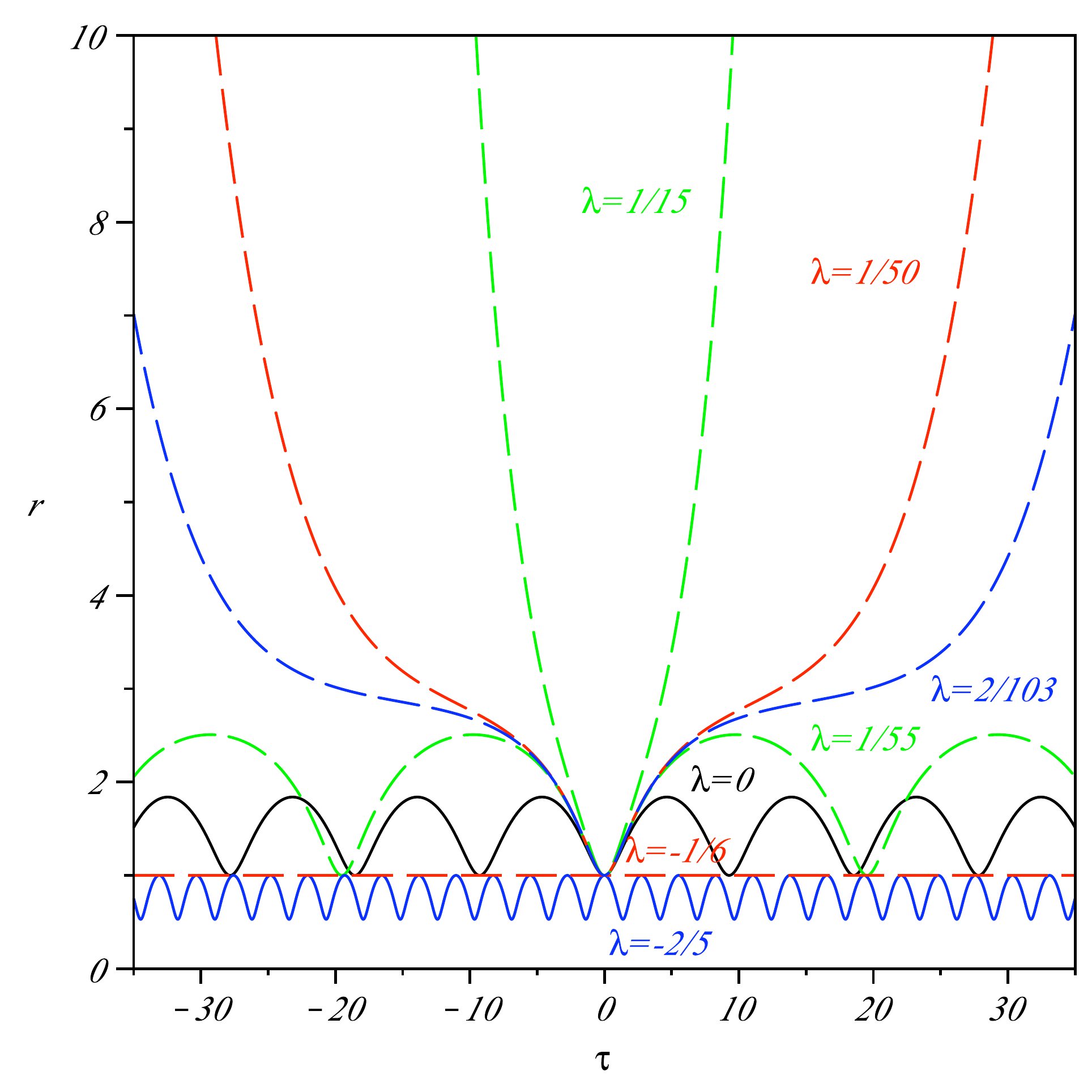}
		\hspace{-1cm}\caption{$\left(w=0, \alpha={\textstyle\frac{1}{2}}\right)$: $\quad$ $r(\tau)$ curves for $\qquad\qquad\qquad$ $\lambda=\{{-\textstyle\frac{2}{5}}, {-\textstyle\frac{1}{6}}, 0, {\textstyle\frac{1}{55}}, {\textstyle\frac{2}{103}}, {\textstyle\frac{1}{50}}, {\textstyle\frac{1}{15}}\}$}
		\label{w_0_a_1_2}
	\vspace{1.5cm}
	\end{minipage}
	\hspace{0.5cm}
	\begin{minipage}[b]{0.5\linewidth}
		\centering
		\includegraphics[scale=0.35]{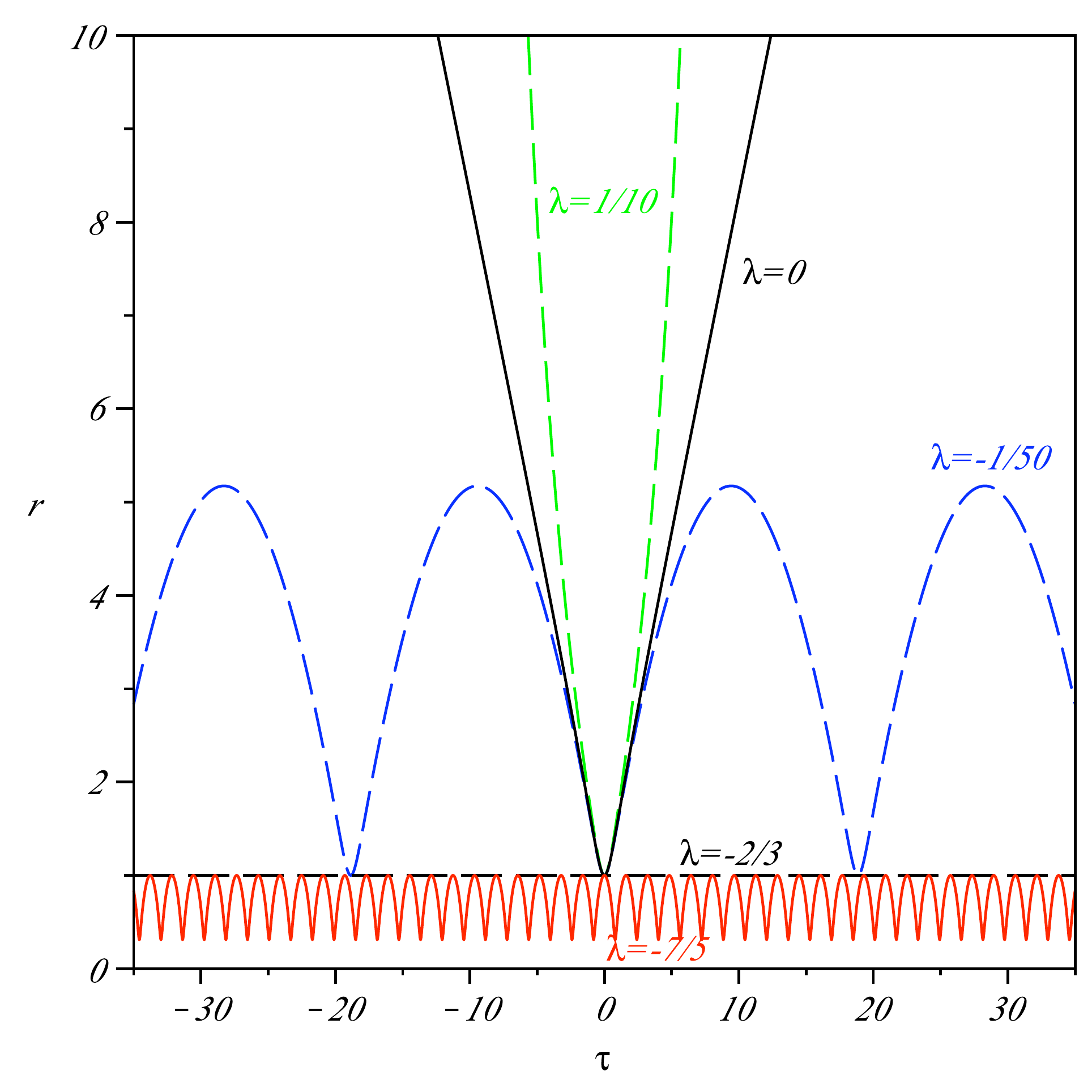}
		\caption{$\left(w={\textstyle\frac{1}{3}}, \alpha=-{\textstyle\frac{1}{2}}\right)$: $\quad$ $r(\tau)$ curves for $\qquad\qquad\qquad$ $\lambda=\{-{\textstyle\frac{7}{5}}, -{\textstyle\frac{2}{3}}, -{\textstyle\frac{1}{50}}, 0, {\textstyle\frac{1}{10}}\}$}
		\label{w_1_3_a_-1_2}
	\vspace{1.5cm}
	\end{minipage}
\end{figure}

\begin{figure}
	\begin{minipage}[b]{0.5\linewidth}
		\centering
		\includegraphics[scale=0.35]{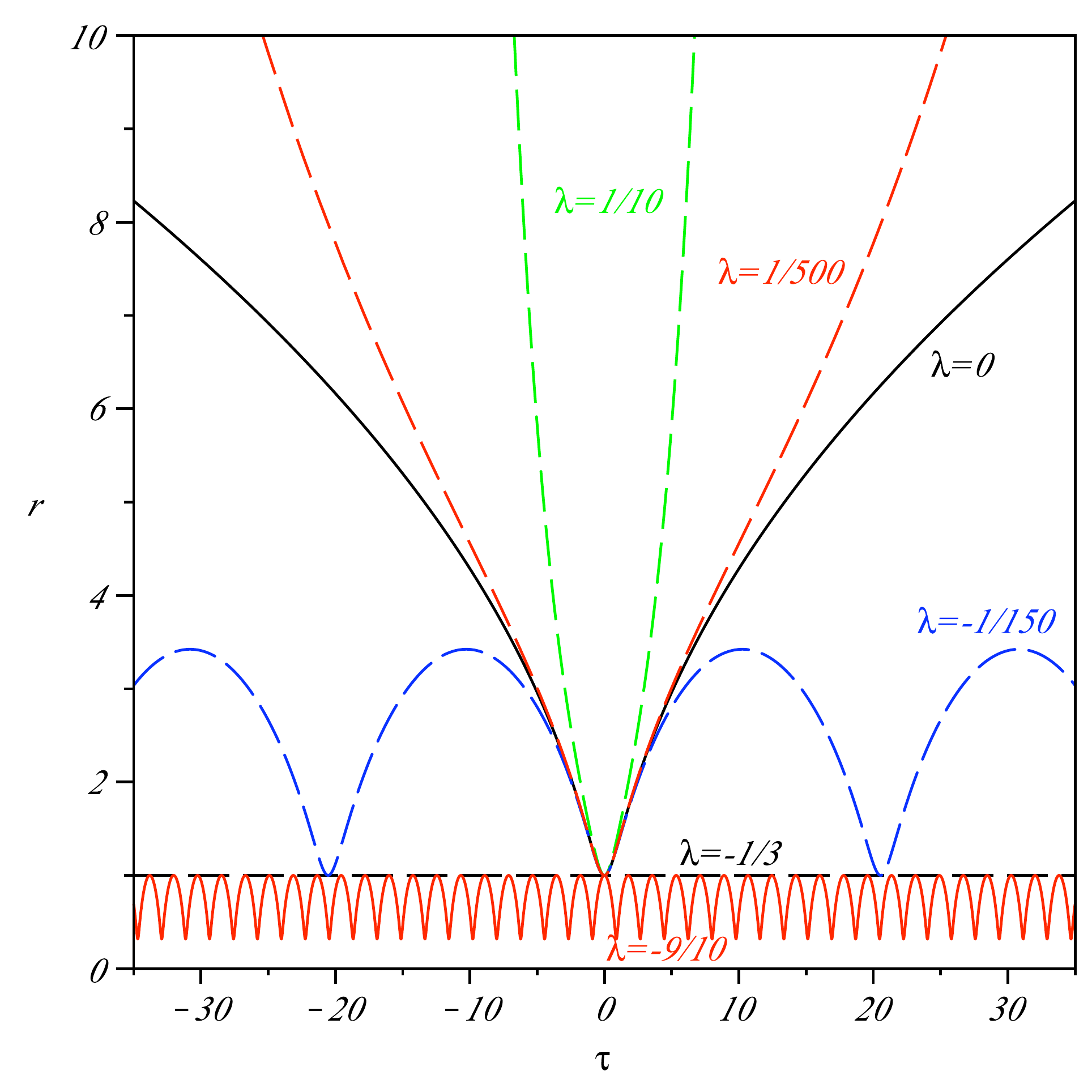}
		\caption{$\left(w={\textstyle\frac{1}{3}}, \alpha=0\right)$: $\quad$ $r(\tau)$ curves for $\qquad\qquad\qquad$ $\lambda=\{-{\textstyle\frac{9}{10}}, -{\textstyle\frac{1}{3}}, -{\textstyle\frac{1}{150}}, 0, {\textstyle\frac{1}{500}}, {\textstyle\frac{1}{10}}\}$}
		\label{w_1_3_a_0}
	\end{minipage}
	\hspace{0.5cm}
	\begin{minipage}[b]{0.5\linewidth}
		\centering
		\includegraphics[scale=0.35]{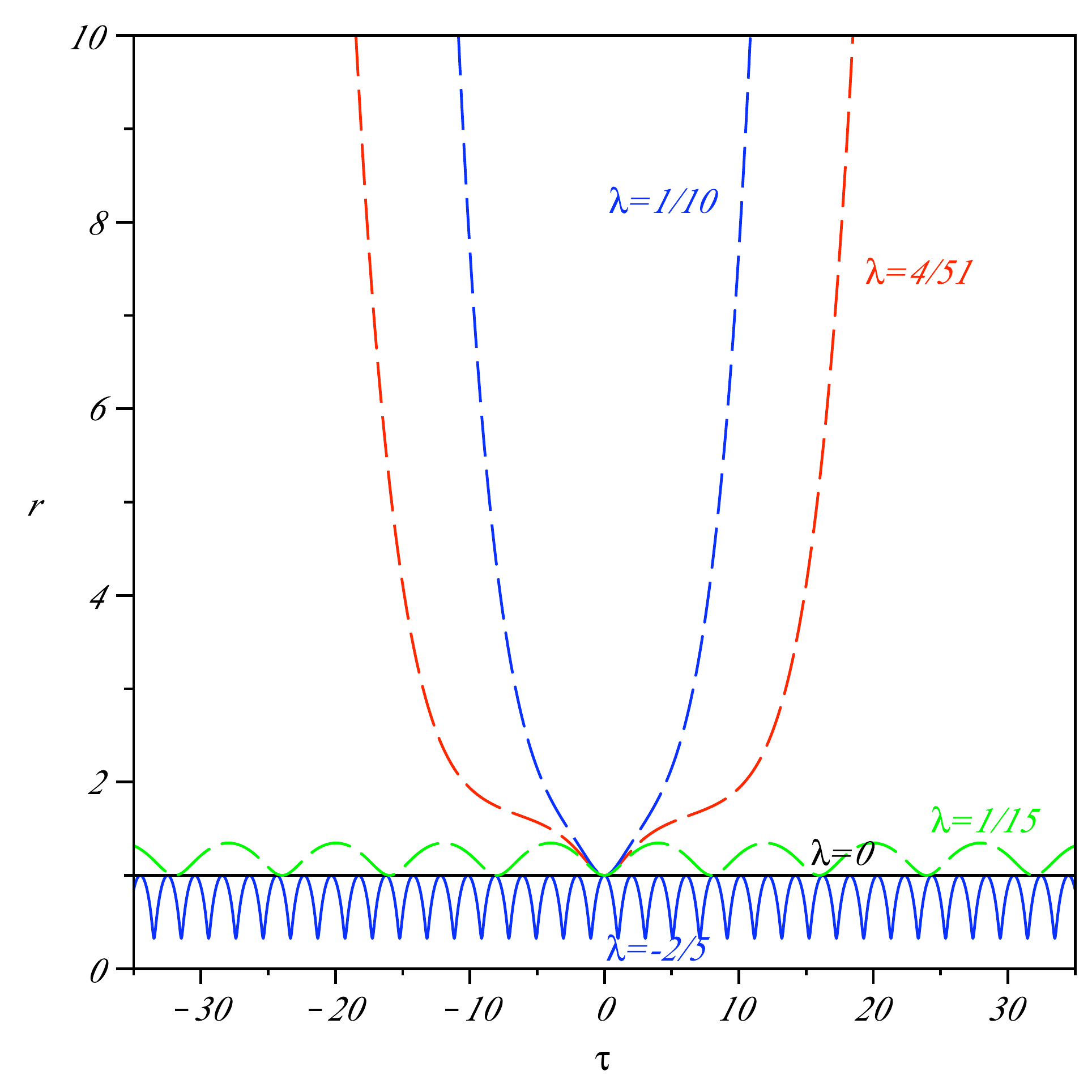}
		\caption{$\left(w={\textstyle\frac{1}{3}}, \alpha={\textstyle\frac{1}{2}}\right)$: $\quad$ $r(\tau)$ curves for $\qquad\qquad\qquad$ $\lambda=\{{-\textstyle\frac{2}{5}}, 0, {\textstyle\frac{1}{15}}, {\textstyle\frac{4}{51}}, {\textstyle\frac{1}{10}}\}$}
		\label{w_1_3_a_1_2}
	\end{minipage}
\end{figure}

The behaviour of the solutions for the scale factor parameter $r(\tau)$ is summarised in $\Tref{summary r solutions lambda}$ below. Explicit numerical solutions in presence of a cosmological constant for particular values of the curvature parameter $\alpha=\{{\textstyle-\frac{1}{2}},0,{\textstyle\frac{1}{2}}\}$ and equation-of-state parameter $w=\{-1,-{\textstyle\frac{1}{3}},0,{\textstyle\frac{1}{3}}\}$ are displayed in $\Fref{w_-1_a_-1_2}$ - $\Fref{w_1_3_a_1_2}$. 

\begin{table}[ht]
\centering
\caption{Behaviour of the solutions $r(w,\alpha,\lambda)$}
\label{summary r solutions lambda}
\begin{tabular}{c|c}
\multicolumn{2}{c}{}\\
\hline\hline
$\lambda$ & $r$\\
\hline
$\lambda>{\textstyle\frac{2}{3}}\alpha+{\textstyle\frac{1}{2}}(w-1)$ \T\B & $1\leq r\leq \infty$ \T\B \\
\hline
$\lambda={\textstyle\frac{2}{3}}\alpha+{\textstyle\frac{1}{2}}(w-1)$ \T\B & $r=1$ \T\B \\
\hline
$\lambda<{\textstyle\frac{2}{3}}\alpha+{\textstyle\frac{1}{2}}(w-1)$ \T\B & $0<r^*\leq r\leq 1$ \T\B \\
\hline
\end{tabular}
\end{table}

\pagebreak

\subsection{Solutions in the absence of a cosmological constant}

\label{curvature}

In the absence of a cosmological constant, the consistency condition at the bounce $\eref{Derivability cond lambda}$ reduces to,
\begin{eqnarray}
s^2-\delta^2=1-\alpha\ .
\label{Derivability cond alpha}
\end{eqnarray}
Using the consistency condition $\eref{Derivability cond alpha}$, the Friedmann $\eref{Friedmann lambda}$ and Raychaudhuri $\eref{Raychaudhuri lambda}$ respectively reduce to,
\begin{eqnarray}
{r^{\prime}}^2=\frac{1}{r^4}\left(r^{3(1-w)}-\alpha\left(r^4-1\right)-1\right)\ ,\label{Friedmann alpha}\\
r^{\prime\prime}=-\frac{1}{r^5}\left(\frac{1+3w}{2}r^{3(1-w)}+2\left(\alpha-1\right)\right)\label{Raychaudhuri alpha}\ .
\end{eqnarray}

As in the presence of a cosmological constant, the behaviour of the solutions can be deduced from the asymptotic behaviour of the expansion rate parameter $r^{\prime}$ and its derivative $r^{\prime\prime}$. The corresponding results concerning limiting values of $r^{\prime}$ and $r^{\prime\prime}$ are obtained by setting $\lambda=0$ in $\eref{i}$, $\eref{ii}$, $\eref{iii}$ and $\eref{iiii}$ . In this simpler case, let us now consider the behaviour of the expansion rate parameter $r^{\prime}$ in the  limit where $r\rightarrow\infty$ and $r\rightarrow 0$.
\begin{enumerate}
\item $w\leq{\textstyle -\frac{1}{3}}$
\begin{equation}
\lim_{r \to \infty}{r^{\prime}}^{2}=\infty\geq 0\ ,
\end{equation}
which implies that the solutions for the scale factor parameter $r(\tau)$ diverge independently of the value of $\alpha$.
\item ${\textstyle -\frac{1}{3}}\leq w\leq 1$
\begin{equation}
\lim_{r \to \infty}{r^{\prime}}^{2}=-\alpha\geq 0\ ,
\end{equation}
which implies that the solutions for the scale factor parameter $r(\tau)$ diverge only for a non-closed spatial geometry (i.e. $\alpha\leq 0$). Hence for a weakly closed spatial geometry (i.e $r^{\prime\prime}>0$ and $0<\alpha<\frac{3}{4}\left(1-w\right)$), the scale factor parameter oscillates between a minimum value $r=1$ and a maximum value $r^*$ defined by $\lim_{r \to r^*}r^{\prime}=0$ according to,
\begin{equation}
1\leq r\leq r^*
\end{equation}
\item $w<1$,
\begin{equation}
\lim_{r \to 0}{r^{\prime}}^{2}=-\infty< 0\ ,
\end{equation}
which does clearly not exist. Hence, the solutions always satisfies $r>0$, which means that there cannot be any singularity. For a strongly closed spatial geometry (i.e $r^{\prime\prime}<0$ and $0<\frac{3}{4}\left(1-w\right)<\alpha<1$), the scale factor parameter oscillates between a maximum value $r=1$ and a minimum value $r^*$ defined by $\lim_{r \to r^*}r^{\prime}=0$ according to,
\begin{equation}
0<r^*\leq r\leq 1
\end{equation}
\end{enumerate}

\begin{figure}[htp]
	\begin{minipage}[b]{0.5\linewidth}
		\centering
		\includegraphics[scale=0.35]{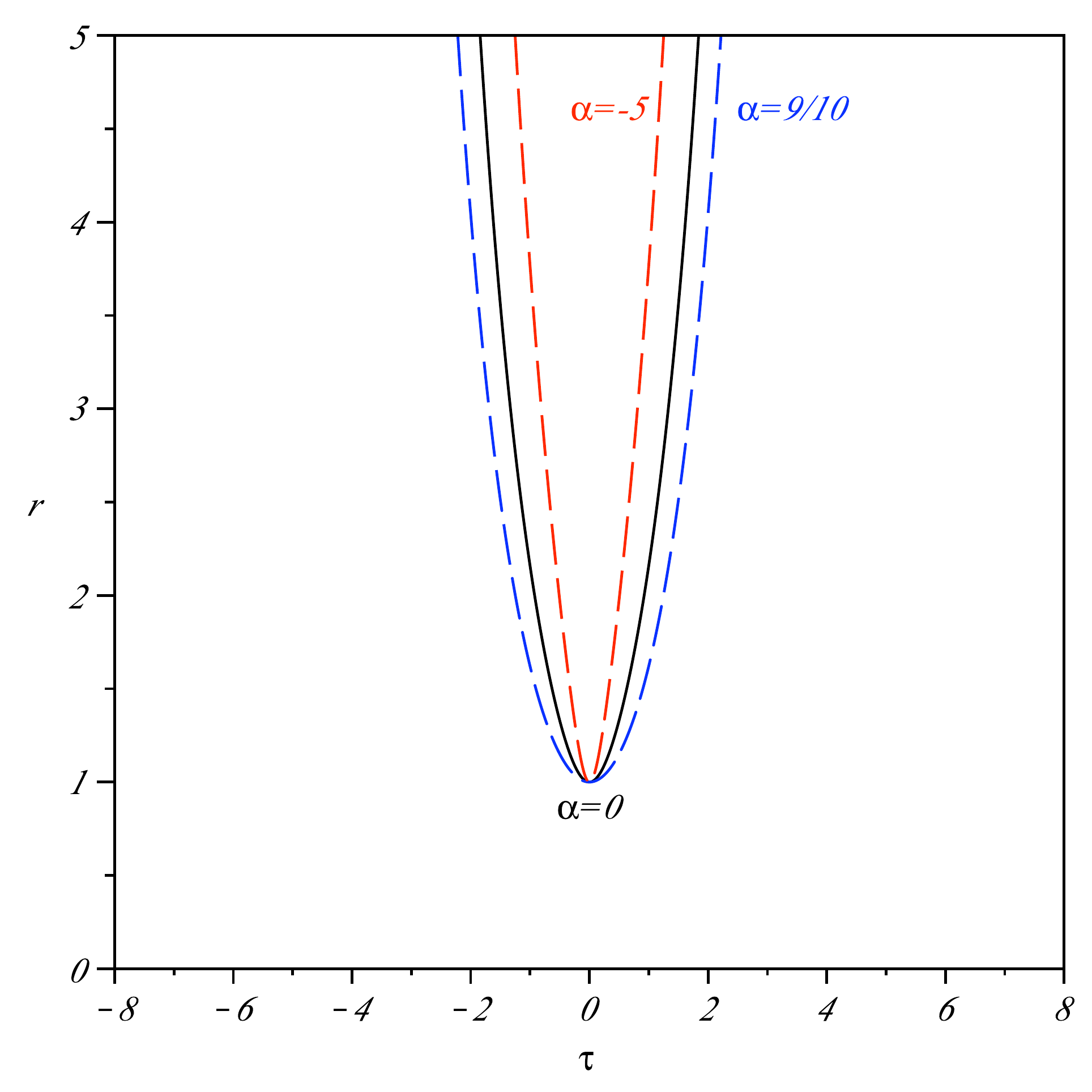}
		\hspace{-1cm}\caption{$\left(w=-1\right)$: $\qquad\quad$   $r(\tau)$ curves for $\alpha=\{-5,0,{\textstyle \frac{9}{10}}\}$}
		\label{w=-1}
	\vspace{1cm}
	\end{minipage}
	\hspace{0.5cm}
	\begin{minipage}[b]{0.5\linewidth}
		\centering
		\includegraphics[scale=0.35]{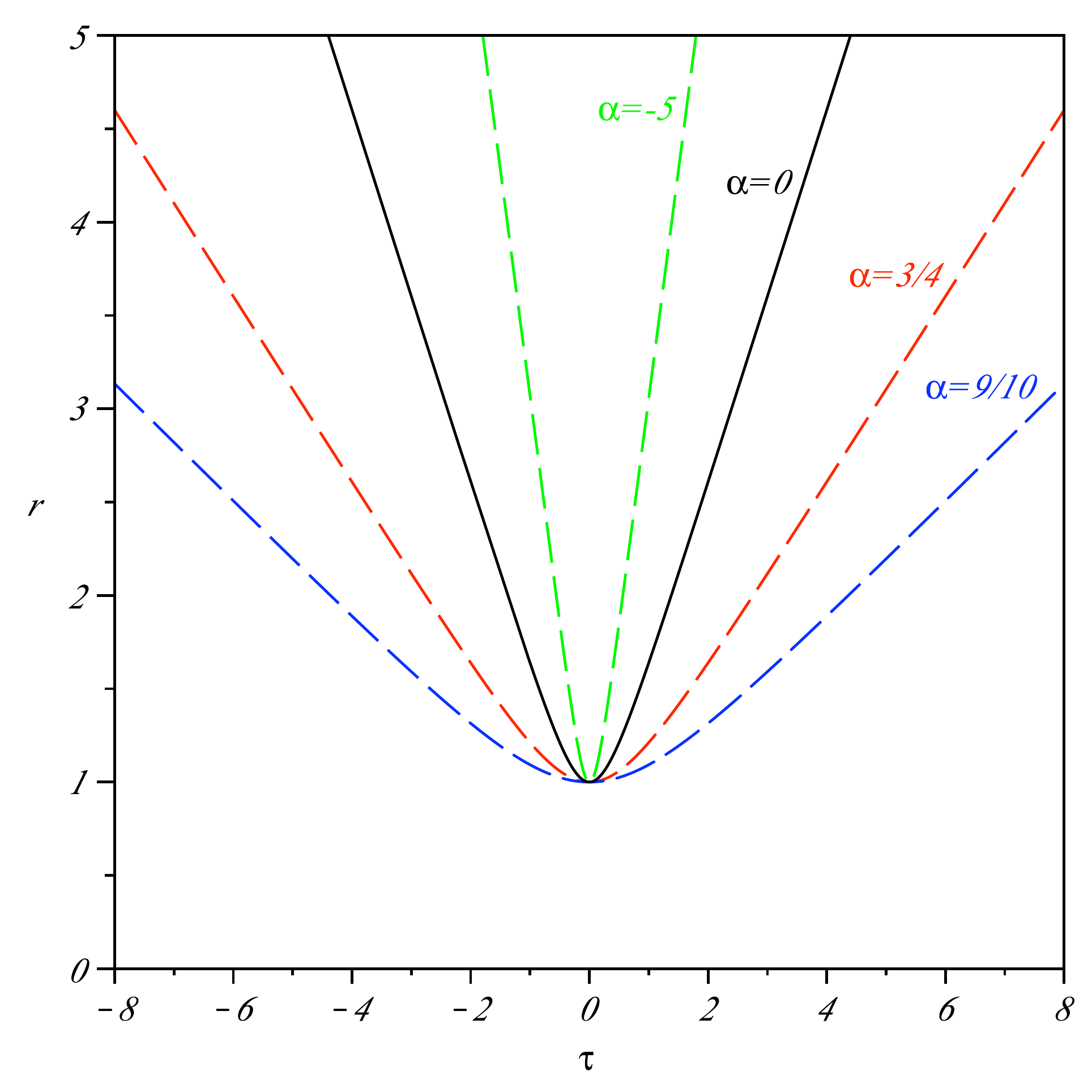}
		\caption{$\left(w={\textstyle -\frac{1}{3}}\right)$: $\qquad\quad$ $r(\tau)$ curves for $\alpha=\{-5,0,{\textstyle \frac{3}{4}},{\textstyle \frac{9}{10}}\}$}
		\label{w=-1/3}
	\vspace{1cm}
	\end{minipage}
	\begin{minipage}[b]{0.5\linewidth}
		\centering
		\includegraphics[scale=0.35]{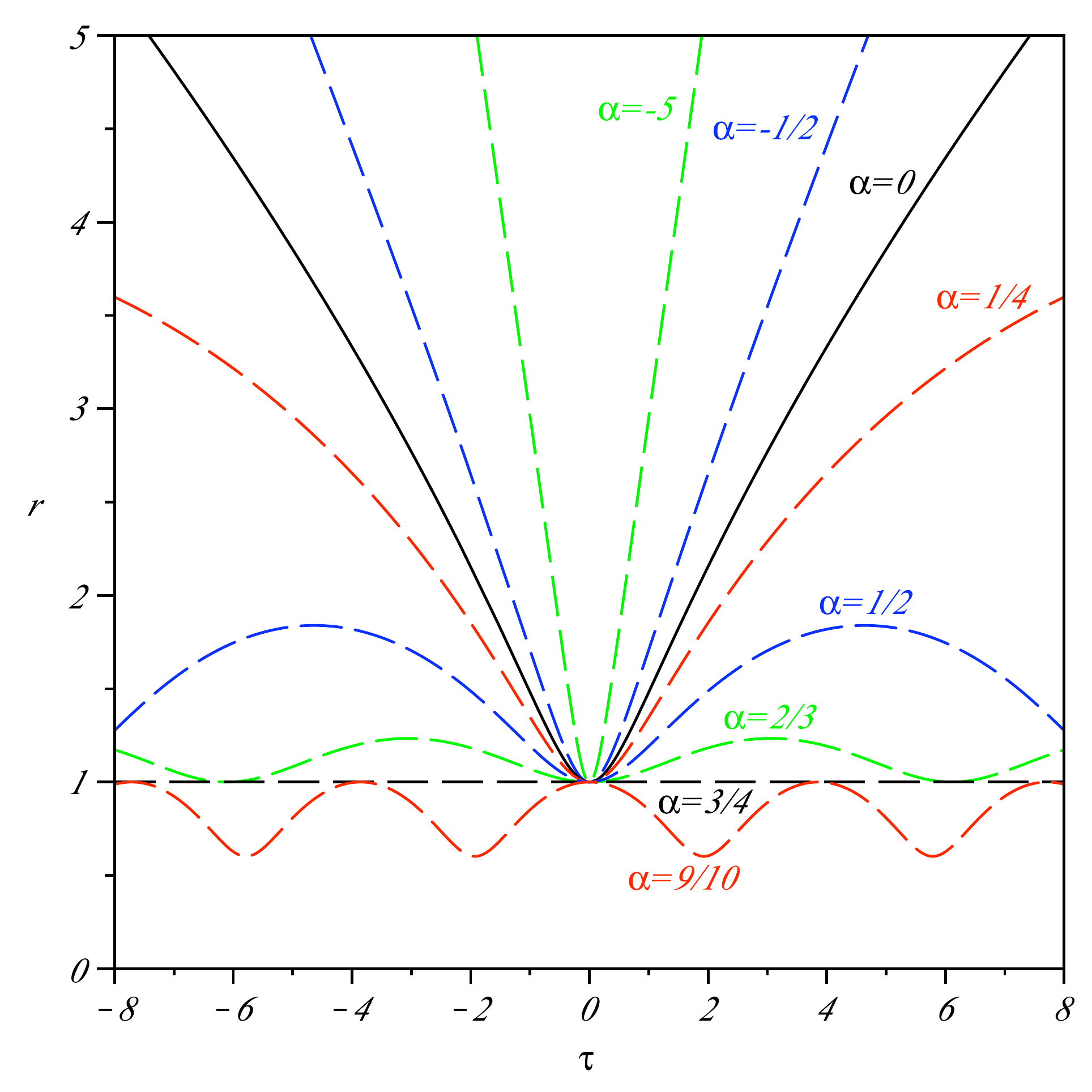}
		\caption{$\left(w=0\right)$: $\quad$ $r(\tau)$ curves for $\qquad$ $\alpha=\{-5,{\textstyle -\frac{1}{2}},0,{\textstyle \frac{1}{4}},{\textstyle \frac{1}{2}},{\textstyle \frac{2}{3}},{\textstyle \frac{3}{4}},{\textstyle \frac{9}{10}}\}$}
		\label{w=0}
	\end{minipage}
	\hspace{0.5cm}
	\begin{minipage}[b]{0.5\linewidth}
		\centering
		\includegraphics[scale=0.35]{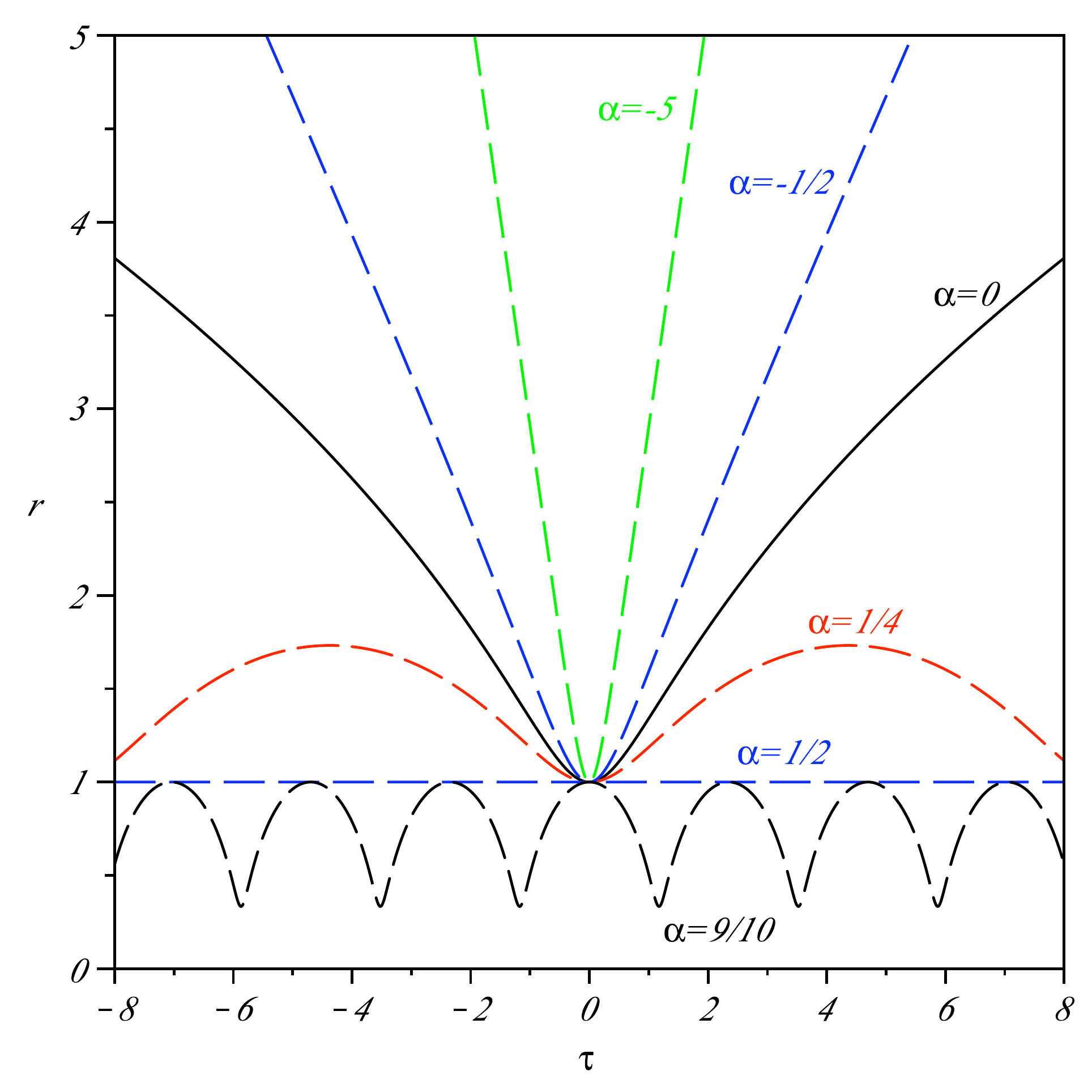}
		\caption{$\left(w={\textstyle \frac{1}{3}}\right)$: $\qquad\qquad$ $r(\tau)$ curves for  $\qquad\qquad\qquad$ $\alpha=\{-5,{\textstyle -\frac{1}{2}},0,{\textstyle \frac{1}{4}},{\textstyle \frac{1}{2}},{\textstyle \frac{9}{10}}\}$}
		\label{w=1/3}
	\end{minipage}
\end{figure}

The behaviour of the solutions for the scale factor parameter $r(\tau)$ is summarised in $\Tref{summary r solutions}$ below. Explicit numerical solutions in presence of curvature (i.e. $\alpha\neq 0$) for $w=\{-1,{\textstyle-\frac{1}{3}},0,{\textstyle\frac{1}{3}}\}$ are displayed in $\Fref{w=-1}$ - $\Fref{w=1/3}$. 

\begin{table}[ht]
\caption{Behaviour of the solutions $r(w,\alpha)$ for $\lambda=0$}
\label{summary r solutions}
\centering
\begin{tabular}{c|c|c}
\multicolumn{3}{c}{}\\
\hline\hline
$w$ & $\alpha$ & $r$\\
\hline
$w\leq{\textstyle-\frac{1}{3}}$ \T\B & $\alpha<1$ \T\B & $1\leq r\leq \infty$ \T\B \\
\hline
\multirow{4}{*}{${\textstyle-\frac{1}{3}}<w<1$} & $\alpha\leq 0$ \T\B& $1\leq r\leq\infty$ \T\B\\ \cline{2-3}
 & $0<\alpha<\frac{3}{4}\left(1-w\right)<1$ \T\B & $1\leq r\leq r^*$ \T\B\\ \cline{2-3}
 & $0<\alpha=\frac{3}{4}\left(1-w\right)<1$ \T\B & $r=1$ \T\B\\ \cline{2-3}
 & $0<\frac{3}{4}\left(1-w\right)<\alpha<1$ \T\B & $0<r^*\leq r\leq 1$ \T\B\\ \cline{2-3}
\hline
\end{tabular}
\end{table}

\section{Dynamical evolution of spatially-flat models with zero cosmological constant}
\label{homogeneous solutions}

In this section, we restrict our study to models with a vanishing spatial curvature and cosmological constant (i.e. $^{*}\mathcal{R}=\Lambda=0$) and find explicit solutions for the time evolution of the scale factor. The reason for choosing this particular class of models is because they admit analytical solutions. The dynamics of a homogeneous and anisotropic Weyssenhoff fluid in a spatially-flat model in the absence of a cosmological constant can be solved explicitly by determining the asymptotic behaviour of the time evolution of the scale factor for particular values of the equation of state parameter.

It is worth mentioning that exact bouncing solutions for spatially-flat cosmological models based on metric affine gravity theories (MAG), which include the solutions for the class of models based on a Weyssenhoff fluid, have previously been discussed by Stachowiak and Szydlowski $\cite{Stachowiak:2007}$. 

In the absence of curvature and a cosmological constant, the consistency condition at the bounce $\eref{Derivability cond lambda}$ reduces to,
\begin{eqnarray}
s^2-\delta^2=1\ .
\label{Derivability cond delta}
\end{eqnarray}
Using the consistency condition at the bounce $\eref{Derivability cond delta}$ the Friedmann $\eref{Friedmann lambda}$ and Raychaudhuri $\eref{Raychaudhuri lambda}$ equations can be rewritten in terms of the dimensionless parameters according to,
\begin{eqnarray}
\left(\frac{r^{\prime}}{r}\right)^2=r^{-3(1+w)}-r^{-6}\ ,\label{Friedmann equ}\\
\frac{r^{\prime\prime}}{r}=-\frac{1+3w}{2}r^{-3(1+w)}+2r^{-6}\ .\label{Raychaudhuri equ}
\end{eqnarray}   

To obtain the explicit time evolution for the scale factor parameter $r$, the Friedmann $\eref{Friedmann equ}$ or Raychaudhuri $\eref{Friedmann equ}$ equations have to be integrated. In order to obtain an analytical result, it is easier to integrate the Friedmann equation $\eref{Friedmann equ}$ according to,

\begin{eqnarray}
\int\frac{r^2dr}{\sqrt{r^{3(1-w)}-1}}=\int d|\tau|\ .\label{Friedmann integrated param}
\end{eqnarray}

The solution of this integral relation depends critically on the value of the equation of state parameter $w$. We will consider six special cases given respectively by $w=\{-1,-\textstyle\frac{1}{3},0,\textstyle\frac{1}{3},1,2\}$, which all admit analytical solutions to $\eref{Friedmann integrated param}$. The last two solutions (i.e. $w=1,2$) are physically unacceptable $\eref{upper equ state}$ but mathematically interesting solutions.

We first note, however, that in the limit where the model approaches the bounce ($r\rightarrow 1$), the asymptotic solution for the scale factor parameter has the quadratic form,
\begin{eqnarray}
r(\tau)=\left(1+\frac{3}{4}(1-w)\tau^2\right)\ ,\label{asympt bounce}
\end{eqnarray}
for any equation-of-state parameter $w$.

Moreover, in the limit where the model is sufficiently far away from the bounce ($r\gg 1$), the asymptotic solution for the scale factor parameter is given by,
\begin{eqnarray}
r(\tau)=\ \mathrm{exp}\left(|\tau|\right)\ ,\label{late -1}
\end{eqnarray}
for an equation-of-state parameter $w=-1$, and evolves according to,
\begin{eqnarray}
r(\tau)=\left(\frac{3}{2}\left(1+w\right)\tau\right)^{\frac{2}{3(1+w)}}\ ,\label{asympt GR}
\end{eqnarray}
for an equation-of-state parameter $w$ satisfying $-1<w<1$. Hence, for a positively defined cosmic time parameter ($\tau>0$), the asymptotic solutions for the scale factor parameter at late times, ($r\gg 1$), have the same time dependence as the solutions found within a GR framework. This is due to the fact that the spin contributions can be neglected at late times, which implies that the evolution of an effective Weyssenhoff fluid asymptotically reduces to a perfect fluid in GR at late times. 

\subsection{$w=-1$ case}

A fluid with an equation-of-state parameter $w=-1$ behaves like a cosmological constant. By solving the integrated Friedmann equation $\eref{Friedmann integrated param}$ for such an equation-of-state parameter, the symmetric evolution of the scale factor parameter with respect to the cosmic time parameter is found to be,
\begin{eqnarray}
r=\left(\mathrm{cosh}\left(3\tau\right)\right)^{1/3}\ .\label{implicit func iso}	
\end{eqnarray}
For a Weyssenhoff fluid satisfying such an equation-of-state parameter, the symmetric temporal curvature of the scale factor parameter $\ddot{r}(\tau)$ is positively defined at all times. Hence, for a positively defined cosmic time parameter ($\tau>0$), the model inflates perpetually. It starts with a power law inflation phase $\eref{asympt bounce}$ and tends towards an exponentially inflating solution at late times $\eref{late -1}$.

\subsection{$w=-{\textstyle\frac{1}{3}}$ case}

A fluid with $w=-{\textstyle\frac{1}{3}}$ behaves like a macroscopic fluid made of cosmic strings. This result has been established by Vilenkin by performing a spatial averaging over a chaotic distribution of linear strings made of matter fields $\cite{Vilenkin:1981}$. For such an equation-of-state parameter, an implicit relation for the symmetric time evolution of the scale factor parameter is found according to,
\begin{eqnarray}
\fl |\tau|=\frac{1}{r}\sqrt{r^4-1}+\mathfrak{Re}\left(\frac{1}{\sqrt{2}}\mathrm{F}\left(\mathrm{arccos}\left(\frac{1}{r}\right),\frac{1}{\sqrt{2}}\right)-\sqrt{2}\mathrm{E}\left(\mathrm{arccos}\left(\frac{1}{r}\right),\frac{1}{\sqrt{2}}\right)\right)\ ,\label{implicit func -1_3}	
\end{eqnarray}
where $F(\phi,k)$ and $E(\phi,k)$ are the elliptic integral of the first and second kind respectively. As in the previous case, the symmetric temporal curvature of the scale factor parameter $\ddot{r}(\tau)$ is positively defined at all times. For a positively defined cosmic time parameter ($\tau>0$), the scale factor parameter $r(\tau)$ tends asymptotically towards a constant rate of expansion (i.e. $\mathop {\lim }\limits_{\tau \to \infty} \ddot{r}(\tau) = 0$) in this limiting case.

\subsection{$w=0$ case}
A fluid with $w=0$ behaves like dust. The non-singular behaviour of dust with spin was first investigated by Trautman $\cite{Trautman:1973}$ and extended by Kuchowicz $\cite{Kuchowicz:1973}$. The integrated Friedmann equation $\eref{Friedmann integrated param}$ for an isotropic Weyssenhoff dust can be solved exactly. The symmetric evolution of the scale factor parameter with respect to the cosmic time parameter is given by,
\begin{eqnarray}
r=\left(1+\frac{9}{4}\tau^2\right)^{1/3}\ ,\label{inflation dust}		
\end{eqnarray}
which agrees with the result established by Trautman. 

\subsection{$w={\textstyle\frac{1}{3}}$ case}

A fluid with $w={\textstyle\frac{1}{3}}$ behaves like radiation. For such an equation-of-state parameter, an implicit relation for the symmetric time evolution of the scale factor parameter is found according to,
\begin{eqnarray}
|\tau|=\frac{1}{2}\left(r\sqrt{r^2-1}+\mathrm{arccosh}\left(r\right)\right)\ .\label{implicit func 1_3}	
\end{eqnarray}
As in the anisotropic case, the isotropic solution of the scale factor parameter for a relativistic fluid with spin $\eref{implicit func 1_3}$ has a clear physical meaning. It is an interpolation between two limiting solutions, which describe an inflationary $\eref{asympt bounce}$ and a radiation dominated $\eref{asympt GR}$ era respectively.

\subsection{$w=1$ case}

A fluid with $w=1$ behaves like stiff matter. For such an equation-of-state parameter, the derivative of the integrated Friedmann equation $\eref{Friedmann integrated param}$ with respect to the cosmic parameter yields a vanishing rate of expansion, 
\begin{eqnarray}
r^{\prime}=0\ .\label{cond w1 case}	
\end{eqnarray}
The value of the scale factor parameter at the bounce is given by $r(0)=1$. Hence, the trivial solution for the evolution of the scale parameter with respect to the cosmic time parameter is found according to,
\begin{eqnarray}
r=1\ .\label{implicit func 1}	
\end{eqnarray}

\subsection{$w=2$ case}

Finally, a fluid with $w=2$ behaves like ultra stiff matter. A fluid with an equation-of-state parameter $w>1$ is physically unreasonable given that for such a fluid the speed of sound exceeds the speed of light ($c_s>c$). However, such a solution is mathematically interesting because it leads to the presence of singularities. By solving the integrated Friedmann equation $\eref{Friedmann integrated param}$ for an equation-of-state parameter $w=2$, an implicit relation for the symmetric time evolution of the scale factor parameter is found according to,
\begin{eqnarray}
|\tau|=\frac{1}{3}\left(\sqrt{r^3\left(1-r^3\right)}+\mathrm{arctan}\sqrt{r^{-3}-1}\right)\ .\label{implicit func 2}	
\end{eqnarray}

For a Weyssenhoff fluid with an equation-of-state parameter $w=2$, the time-symmetric temporal curvature of the scale factor parameter $\ddot{r}(\tau)$ is negatively defined at all times $\eref{curvature 2}$. To ensure the continuity of the expansion rate $\Theta$ at the bounce, the energy density at the bounce $\rho_0$ has to satisfy $\eref{Friedmann start Cond}$ even if the cosmological solution leads to the presence of singularities. As the absolute value of the cosmic time parameter $|\tau|$ increases, the value of the scale factor parameter decreases before eventually collapsing. From the implicit dynamical relation $\eref{implicit func 2}$, the cosmic time parameter $|\tau_c|$ at the collapse $\--$ defined by a vanishing scale factor parameter $r(|\tau_c|)=0$ $\--$ is found to be,
\begin{eqnarray}
|\tau_c|=\frac{\pi}{6}\ .\label{tau coll}
\end{eqnarray}
The collapse of the scale factor $R\rightarrow 0$ is equivalent to the divergence of the expansion rate $\Theta\rightarrow\infty$. For an equation of state parameter $w=2$, the collapse of the scale factor parameter represents a mathematical singularity for the evolution of the scale factor parameter with respect to the cosmic time parameter given that the rate of expansion diverges at that point, ${\Theta}(\tau_c)=\infty$.

\subsection{Graphic solutions}

\begin{figure}[htbp]
\begin{center}
\includegraphics[width=11cm,height=9cm]{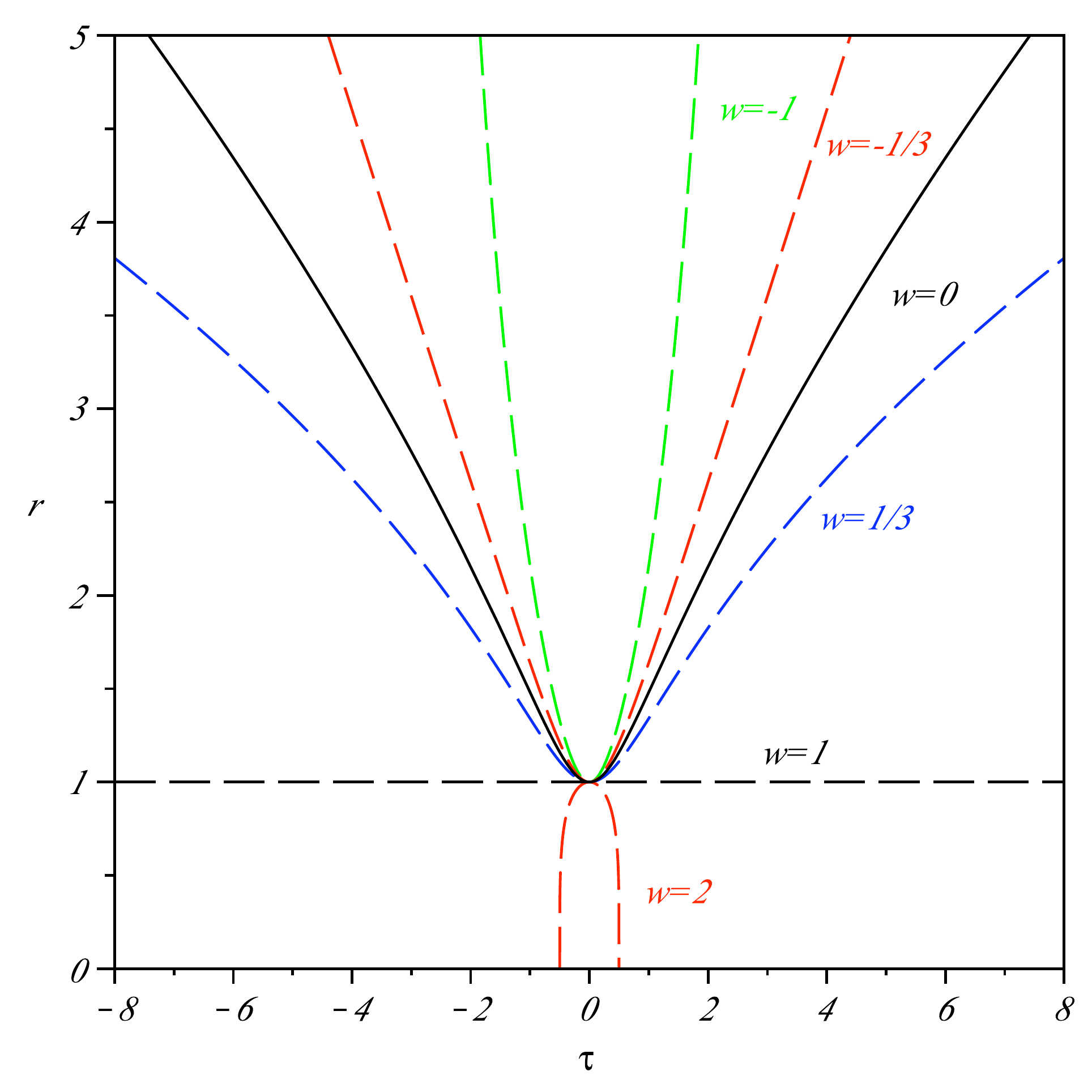}
\caption{Symmetric evolution of the scale factor parameter $r$ with respect to the cosmic time parameter $\tau$ for particular values of the equation-of-state parameter $w=\{ -1, -{\textstyle\frac{1}{3}}, 0, {\textstyle\frac{1}{3}}, 1, 2\}$ for spatially-flat models with zero cosmological constant.\label{scale factor 1}}
\end{center}
\end{figure}

The cosmological constant ($w=-1$), cosmic strings ($w=-{\textstyle\frac{1}{3}}$), dust ($w=0$), radiation ($w={\textstyle\frac{1}{3}}$), stiff matter ($w=1$) and ultra stiff matter ($w=2$) solutions for the evolution of the scale factor parameter with respect the cosmic time parameter $r(\tau)$ are shown in $\Fref{scale factor 1}$.  For a positively defined cosmic time parameter $(\tau>0)$, the inflection point on the graph of $r(\tau)$ $\--$ for the dust and radiation solutions $\--$ corresponds to the end of inflation. The coordinates of this point are $(1.28,1.41)$ for the radiation case and $(1.15,1.59)$ for the dust case.

\section{Conclusions}

We have used the $1+3$ covariant approach to perform a dynamical analysis of an effective homogeneous and irrotational Weyssenhoff fluid. Contrary to the case of a perfect fluid in GR, the effective spin contributions to the fluid dynamics act like centrifugal forces preventing the formation of singularities for isotropic and anisotropic models satisfying the spin-shear constraint $\eref{cond spin shear}$. The temporal evolution of the models is symmetric with respect to $t=0$.

In a cosmological context, the energy density at the bounce state $\rho_0$ has to be sufficiently dense in order to seed large scale structures from primordial quantum fluctuations. For cosmological parameters which are consistent with current cosmological data $\eref{lambda condition}$ $\eref{curvature condition}$, the temporal curvature of scale factor of a Weyssenhoff fluid is positively defined near the bounce $\eref{a}$. However such a fluid is not a suitable candidate for inflation given that the only way to include an inflation phase of about $50-70$ e-folds, is by considering a fluid with a very fine-tuned equation-of-state $\eref{State tunning}$, which does not reduce to the standard cosmological fluid at later times.

It is worth emphasizing that the time evolution of the scale factor of a homogeneous and irrotational Weyssenhoff fluid exhibits eternal oscillations, without any singularities. By contrast, the corresponding solutions obtained for a perfect fluid in GR are cycloids, which do exhibit singularities. Hence, the absence of singularities for a specific range of parameters is a genuinely new feature of cosmological models based on a Weyssenhoff fluid.

\ack

S~D~B thanks the Isaac Newton Studentship and the Sunburst Fund for their support. The authors also thank Reece Heineke for giving an insightful talk entitled ``Inflation via Einstein-Cartan theory", and John M. Stewart for useful discussions.

\appendix

\section{1+3 covariant formalism}

We choose to restrict our scope to a homogeneous and irrotational Weyssenhoff fluid, thus implying a vanishing vorticity, $\omega_{\mu\nu}=0$, and acceleration, $a_{\mu}=0$. To study the dynamics of an such a fluid, we use the $1+3$ covariant approach which has been described in detail in our previous paper $\cite{Brechet:2007a}$, and will be summarised below to clarify our notation. The approach relies on a $1+3$ decomposition of geometric quantities with respect to a timelike velocity field $u^{\mu}$ defining an observer
according to, 
\begin{equation} 
\nabla_{\mu}u_{\nu}=D_{\mu}u_{\nu}={\textstyle\frac{1}{3}}\Theta h_{\mu\nu}+\sigma_{\mu\nu}=\Theta_{\mu\nu}\ ,\label{kinematics}
\end{equation}
where 
\begin{itemize} 
\item $h_{\mu\nu}\equiv g_{\mu\nu}-u_{\mu}u_{\nu}$ is the induced metric
on the orthogonal instantaneous rest-spaces of observers moving with $4$-velocity $u^{\mu}$. 
\item $D_{\mu}u_{\nu}\equiv {h_{\mu}}^{\rho}{h_{\nu}}^{\sigma}\nabla_{\rho}u_{\sigma}$ is the projected covariant derivative of the worldline on the orthogonal instantaneous rest-space.
\item $\Theta\equiv D_{\mu}u^{\mu}$ is the scalar describing the volume rate of expansion of
the fluid (with $H={\textstyle\frac{1}{3}}\Theta$ the Hubble parameter). 
\item $\sigma_{\mu\nu}\equiv D_{\langle\mu}u_{\nu\rangle}$ is the trace-free rate-of-shear tensor describing the rate of distortion of the
matter flow.
\item $\Theta_{\mu\nu}$ is the symmetric fluid evolution tensor describing the rate of expansion and distortion of the fluid.
\end{itemize}
It is useful to introduce another scalar quantity, namely the rate of shear scalar defined as,
\begin{eqnarray}
\sigma^2={\textstyle\frac{1}{2}}\sigma_{\mu\nu}\sigma^{\mu\nu}\geq 0\ .\label{shear scalar}
\end{eqnarray}
Moreover, we define two projected covariant derivatives which are the time projected covariant derivative along the worldline (denoted $\mathbf{\dot{}}\ $) and the orthogonally projected covariant derivative (denoted $D_{\mu}$). For any general tensor $T^{\mu\dots}_{\ \ \ \ \nu\dots}$, these are respectively defined as 
\begin{eqnarray}
\dot{T}^{\mu\dots}_{\ \ \ \ \nu\dots} &\equiv u^{\lambda}\nabla_{\lambda}T^{\mu\dots}_{\ \ \ \ \nu\dots}\ ,\\
D_{\lambda}T^{\mu\dots}_{\ \ \ \ \nu\dots} &\equiv h^{\epsilon}_{\ \lambda}h^{\mu}_{\ \rho}\dots
h^{\sigma}_{\ \nu}\dots \nabla_{\epsilon}T^{\rho\dots}_{\ \ \ \sigma\dots}\ . 
\end{eqnarray} 
Furthermore, the dynamics is determined by projected tensors that are orthogonal to $u^{\mu}$ on every index. The angle brackets are used to denote respectively the orthogonally projected
symmetric trace-free part $(\mathrm{PSTF})$ of rank-$2$ tensors and their time derivative along the worldline according to, \begin{eqnarray}
T^{\langle \mu\nu\rangle} &= \left(h^{(\mu\vphantom)}_{\
\ \rho}h^{\vphantom(\nu)}_{\ \ \sigma}- {\textstyle\frac{1}{3}}h^{\mu\nu}h_{\rho\sigma}\right)T^{\rho\sigma}\\
\dot{T}^{\langle \mu\nu\rangle} &=
\left(h^{(\mu\vphantom)}_{\ \ \rho}h^{\vphantom(\nu)}_{\ \ \sigma}-
{\textstyle\frac{1}{3}}h^{\mu\nu}h_{\rho\sigma}\right)\dot{T}^{\rho\sigma}
\ .\label{SE tensor} 
\end{eqnarray}
The orthogonal projection of the covariant time derivative of a general tensor $T^{\mu\dots}_{\ \ \ \nu\dots}$ is denoted by, 
\begin{eqnarray} 
\left(T^{\mu\dots}_{\ \ \ \nu\dots}\right)^{\cdot}_{\bot}\equiv h^{\mu}_{\ \rho}\dots h^{\sigma}_{\ \nu}\dots u^{\lambda}\nabla_{\lambda}T^{\rho\dots}_{\ \ \ \sigma\dots}\ .\label{Antisym eq} 
\end{eqnarray}

\end{document}